\documentclass[pra,aps,amsmath,amssymb,twocolumn]{revtex4}
\usepackage{graphicx}
\usepackage{amsmath}
\usepackage[tight]{subfigure}

\newcommand{\quot}[1]{{``#1''}}

\newcommand{\sw}{{{\rm sw}}}  
\newcommand{\swrev}{{{\rm swrev}}}  
\newcommand{\swirr}{{{\rm swirr}}}
\newcommand{\DE}[2]{{(\Delta E^{#1})^{\sw}_{#2_1#2_2}}} 
\newcommand{\DErev}[2]{{(\Delta E^{#1})^{\swrev}_{#2_1#2_2}}} 
\newcommand{\DEirr}[2]{{(\Delta E^{#1})^{\swirr}_{#2_1#2_2}}} 
\newcommand{\DErevzero}[2]{{(\Delta E^{#1})^{\swrev}_{#2_1#2_0}}} 
\newcommand{\DErevinv}[2]{{(\Delta E^{#1})^{\swrev}_{#2_2#2_1}}} 
\newcommand{\CA}{{\textsl{\textsf{C}}^A}}
\newcommand{\FA}{{\textbf{F}^{A}_e}}
\newcommand{\FAt}{{\textbf{F}^{A}_{e,t}}}
\newcommand{\FBt}{{\textbf{F}^{B}_{e,t}}}
\newcommand{\FABt}{{\textbf{F}^{AB}_{e,t}}}
\newcommand{\FI}{\textbf{F}^{I}_e}
\newcommand{\FAB}{\textbf{F}^{AB}_e}
\newcommand{\FO}{\textbf{F}^O_e}
\newcommand{\FOQ}{\textbf{F}^{OQ}_e}

\newcommand{\nA}[1]{{\pmb{n}^A_{#1}}}
\newcommand{\nO}[1]{{\pmb{n}^O_{#1}}}
\newcommand{\betaO}[1]{{\pmb{\beta}^O_{#1}}}

\newcommand{\RA}[1]{{\pmb{\textsl{\textsf{R}}}^A_{#1}}}

\newcommand{\spazio}{{\vskip 2mm\noindent}}

\begin{document}

\bibliographystyle{plainnat}

\title{Rigorous and General Definition of Thermodynamic Entropy}

%



\title{Rigorous and General Definition of Thermodynamic Entropy} 

\author{Gian Paolo Beretta}
\affiliation{Universit\`{a} di Brescia,  Via Branze 38, Brescia,
Italy}

\author{Enzo Zanchini} \affiliation{Universit\`{a} di
Bologna, Viale Risorgimento 2, Bologna, Italy}

%
%


\begin{abstract}

The physical foundations of a variety of emerging technologies ---
ranging from the applications of quantum entanglement in quantum
information to the applications of nonequilibrium bulk and
interface phenomena in microfluidics, biology, materials science,
energy engineering, etc. --- require understanding thermodynamic
entropy beyond the equilibrium realm of its traditional
definition. This paper presents a rigorous logical scheme that
provides a generalized definition of entropy free of the usual
unnecessary assumptions which constrain the theory to the
equilibrium domain. The scheme is based on carefully worded
operative definitions for all the fundamental concepts employed,
including those of system, property, state, isolated system,
environment, process, separable system, system uncorrelated from
its environment, and parameters of a system. The treatment
considers also systems with movable internal walls and/or
semipermeable walls, with chemical reactions  and/or external
force fields, and with small numbers of particles. The definition
of reversible process is revised by introducing the new concept of
scenario. The definition of entropy involves neither the concept
of heat nor that of quasistatic process; it applies to both
equilibrium and nonequilibrium states. The role of correlations on
the domain of definition and on the additivity of energy and
entropy is discussed: it is proved that energy is defined and
additive for all separable systems, while entropy is defined and
additive only for separable systems uncorrelated from their
environment; decorrelation entropy is defined. The definitions of
energy and entropy are extended rigorously to open systems.
Finally, to complete the discussion, the existence of the
fundamental relation for stable equilibrium states is proved, in
our context, for both closed and open systems.
\end{abstract}

\maketitle

\section{Introduction}

Thermodynamics and Quantum Theory are among the few sciences
involving fundamental concepts and universal content that are
controversial and have been so since their birth, and yet continue
to unveil new possible applications and to inspire new theoretical
unification. The basic issues in Thermodynamics have been, and to
a certain extent still are: the range of validity and the very
formulation of the Second Law of Thermodynamics, the meaning and
the definition of entropy, the origin of irreversibility, and the
unification with Quantum Theory
  \citep{HB2008}. The basic issues with
Quantum Theory have been, and to a certain extent still are: the
meaning of complementarity and in particular the wave-particle
duality,  understanding the many faces of the many wonderful
experimental and theoretical results on entanglement, and the
unification with Thermodynamics \citep{Horodecki2001}.

Entropy  has a central role in this situation.  It is astonishing
that after over 140 years since the term entropy has been first
coined by Clausius \citep{ref3}, there is still so much discussion
and controversy about it, not to say confusion. Two recent
conferences, both held in October 2007, provide a state-of-the-art
scenario revealing an unsettled and hard to settle field: one,
entitled \emph{Meeting the entropy challenge} \citep{Meeting},
focused on the many physical aspects (statistical mechanics,
quantum theory, cosmology, biology, energy engineering), the
other, entitled \emph{Facets of entropy} \citep{Facets}, on the
many different mathematical concepts that in different fields
(information theory, communication theory, statistics, economy,
social sciences, optimization theory, statistical mechanics) have
all been termed \emph{entropy} on the basis of some analogy of
behavior  with the \emph{thermodynamic entropy}.

Following the well-known Statistical Mechanics and Information
Theory interpretations of thermodynamic entropy, the term
\emph{entropy} is used in many different contexts wherever the
relevant \emph{state description} is in terms of a probability
distribution over some set of possible events which characterize
the  \emph{system description}. Depending on the context, such
events may be \emph{microstates}, or \emph{eigenstates}, or
\emph{configurations}, or \emph{trajectories}, or
\emph{transitions}, or \emph{mutations}, and so on. Given such a
probabilistic description, the term entropy is used for some
functional of the probabilities chosen as a quantifier of their
\emph{spread} according to some reasonable set of defining axioms
\citep{Lieb}. In this sense, the use of a common name for all the
possible different state functionals that share such broad
defining features, may have some unifying advantage from a broad
conceptual point of view, for example it may suggest analogies and
inter-breeding developments between very different fields of
research sharing similar probabilistic descriptions.

However, from the physics point of view, entropy --- the
\emph{thermodynamic entropy} --- is a single definite property of
every well-defined material system that can be measured in every
laboratory by means of standard measurement procedures. Entropy is
a property of paramount practical importance, because it turns out
\citep{GB2005} to be monotonically related to the difference
$E-\Psi$ between the energy $E$ of the system, above the
lowest-energy state, and the adiabatic availability $\Psi$ of the
system, \emph{i.e.}, the maximum work the system can do in a
process which produces no other external effects. It is therefore
very important that whenever we talk or make inferences about
physical (\emph{i.e.}, thermodynamic) entropy, we first agree on a
precise definition.

In our opinion, one of the most rigorous and general axiomatic
definitions of thermodynamic entropy available in the literature
is that given in \citep{GB2005}, which extends to the
nonequilibrium domain one of the best traditional treatments
available in the literature, namely that presented by Fermi
\citep{Fermi}.

In this paper, the treatment presented in \citep{GB2005} is
assumed as a starting point and the following improvements are
introduced. The basic definitions of system, state, isolated
system, environment, process, separable system, and parameters of
a system are deepened, by developing a logical scheme outlined in
 \citep{Zanchini:1988,Zanchini:1992}. Operative
and general definitions of these concepts are presented, which are
valid also in the presence of internal semipermeable walls and
reaction mechanisms. The treatment of \citep{GB2005} is
simplified, by identifying the minimal set of definitions,
assumptions and theorems which yield the definition of entropy and
the principle of entropy non-decrease. In view of the important
role of entanglement in the ongoing and growing interplay between
Quantum Theory and Thermodynamics, the effects of correlations on
the additivity of energy and entropy are discussed and clarified.
Moreover, the definition of a reversible process is given with
reference to a given \emph{scenario}; the latter is the largest
isolated system whose subsystems are available for interaction,
for the class of processes under exam.

Without introducing the quantum formalism, the approach is
nevertheless compatible with it (and indeed, it was meant to be
so, see, \emph{e.g.}, \citet{HG,QTa,QTb,QT2,QT3,QT4,QT5}); it is
therefore suitable to provide a basic logical framework for the
recent scientific revival of thermodynamics in Quantum Theory
[quantum heat engines
\citep{quantumheatengines,quantumheatengines2}, quantum Maxwell
demons \citep{demon,demon2,demon3}, quantum erasers
\citep{erasers,erasers2}, etc.] as well as for the recent quest
for quantum mechanical explanations of irreversibility [see,
\emph{e.g.}, \citet{recent,recent2,HB2008,recent3}].

The paper is organized as follows. In Section \ref{traditional} we
discuss the drawbacks of the traditional definitions of entropy.
In Section \ref{basic} we introduce and discuss a full set of
basic definitions, such as those of system, state, process, etc.
that  form the necessary unambiguous background on which to build
our treatment. In Section \ref{first} we introduce the statement
of the First Law and the definition of energy. In Section
\ref{second} we introduce and discuss the statement of the Second
Law and, through the proof of three important theorems, we build
up the definition of entropy. In Section \ref{fundamental} we
briefly complete the discussion by proving in our context the
existence of the fundamental relation for the stable equilibrium
states and by defining temperature, pressure, and other
generalized forces. In Section \ref{open} we extend our
definitions of energy and entropy to the model of an open system.
In Section \ref{fundamental-open} we prove the existence of the
fundamental relation for the stable equilibrium states of an open
system. In Section \ref{concl} we draw our conclusions and, in
particular, we note that nowhere in our construction we use or
need to define the concept of \emph{heat}.

\section{\label{traditional}Drawbacks of the traditional
definitions of entropy}

In traditional expositions of thermodynamics, entropy is defined
in terms of the concept of heat, which in turn is introduced at
the outset of the logical development in terms of heuristic
illustrations based on mechanics. For example, in his lectures on
physics, Feynman \citep{1} describes heat as one of several
different forms of energy related to the jiggling motion of
particles stuck together and tagging along with each other
(pp.~1-3 and 4-2), a form of energy which really is just kinetic
energy --- internal motion (p.~4-6), and is measured by the random
motions of the atoms (p.~10-8).  Tisza \citep{2} argues that such
slogans as \quot{heat is motion}, in spite of their fuzzy meaning,
convey intuitive images of pedagogical and heuristic value.

 There are at least three problems with these
illustrations. First, work and heat are not stored in a system.
Each is a mode of transfer of energy from one system to another.
Second, concepts of mechanics are used to justify and make
plausible a notion --- that of heat --- which is beyond the realm
of mechanics; although at a first exposure one might find the idea
of \emph{heat as motion} harmless, and even natural, the situation
changes drastically when the notion of heat is used to define
entropy, and the logical loop is completed when entropy is shown
to imply a host of results about energy availability that contrast
with mechanics. Third, and perhaps more important, heat is a mode
of energy (and entropy) transfer between systems that are very
close to thermodynamic equilibrium and, therefore, any definition
of entropy based on heat is bound to be valid only at
thermodynamic equilibrium.

The first problem is addressed in some expositions. Landau and
Lifshitz  \citep{3} define heat as the part of an energy change of
a body that is not due to work done on it. Guggenheim \citep{4}
defines heat as an exchange of energy that differs from work and
is determined by a temperature difference. Keenan \citep{5}
defines heat as the energy transferred from one system to a second
system at lower temperature, by virtue of the temperature
difference, when the two are brought into communication.  Similar
definitions are adopted in most other notable textbooks that are
too many to list.

None of these definitions, however, addresses the basic problem.
The existence of exchanges of energy that differ from work is not
granted by mechanics. Rather, it is one of the striking results of
thermodynamics, namely, of the existence of entropy as a property
of matter.  As pointed out by Hatsopoulos and Keenan \citep{6},
without the Second Law heat and work would be indistinguishable;
moreover, the most general kind of interaction between two systems
which are very far from equilibrium is neither a heat nor a work
interaction. Following Guggenheim it would be possible to state a
rigorous definition of heat, with reference to a very special kind
of interaction between two systems, and to employ the concept of
heat in the definition of entropy \citep{4}. However, Gyftopoulos
and Beretta \citep{GB2005} have shown that the concept of heat is
unnecessarily restrictive for the definition of entropy, as it
would confine it to the equilibrium domain.  Therefore, in
agreement with their approach, we will present and discuss a
definition of entropy where the concept of heat is not employed.

Other problems are present in most treatments of the definition of
entropy available in the literature:
\begin{enumerate}
    \item many basic concepts, such as those of system, state, property,
isolated system, environment of a system, adiabatic process are
not defined rigorously;
    \item on account of unnecessary assumptions
(such as, the use of the concept of quasistatic process), the
definition holds only for stable equilibrium states
\citep{Callen}, or for systems which are in local thermodynamic
equilibrium \citep{Fermi};
    \item in the traditional logical scheme
    \citep{2,3,4,5,6,Callen,Fermi}, some proofs are incomplete.
\end{enumerate}
To illustrate the third point, which is not well known, let us
refer to the definition in \citep{Fermi}, which we consider one of
the best traditional treatments available in the literature. In
order to define the thermodynamic temperature, Fermi considers a
reversible cyclic engine which absorbs a quantity of heat $Q_2$
from a source at (empirical) temperature $T_2$ and supplies a
quantity of heat $Q_1$ to a source at (empirical) temperature
$T_1$. He states that if the engine performs $n$ cycles, the
quantity of heat subtracted from the first source is $n\, Q_2$ and
the quantity of heat supplied to the second source is $n\, Q_1$.
Thus, Fermi assumes implicitly that the quantity of heat exchanged
in a cycle between a source and a reversible cyclic engine is
independent of the initial state of the source. In our treatment,
instead, a similar statement is made explicit, and proved.

\section{\label{basic}BASIC DEFINITIONS}

\noindent \textbf{Level of description, constituents, amounts of
constituents, deeper level of description}. We will call
\emph{level of description} a class of physical models whereby all
that can be said about  the matter contained in a given region of
space \textsf{\emph{R}}, at a time instant $t$, can be described
by assuming that the matter consists of a set of elementary
building blocks, that we call \emph{constituents}, immersed in the
electromagnetic field. Examples of constituents are: atoms,
molecules, ions, protons, neutrons, electrons. Constituents may
combine and/or transform into other constituents according to a
set of model-specific \emph{reaction mechanisms}.

For instance, at the \emph{chemical level of description} the
constituents  are the different chemical species, i.e.,  atoms,
molecules, and ions; at the \emph{atomic level of description} the
constituents are the atomic nuclei and the electrons; at the
\emph{nuclear level of description} they are the protons, the
neutrons, and the electrons.

The particle-like nature of the constituents  implies that a
counting measurement procedure is always defined and, when
performed in a region of space delimited by impermeable walls, it
is \emph{quantized} in the sense that the measurement outcome is
always an integer number, that we call the \emph{number of
particles}. If the counting is selective for the $i$-th type of
constituent only, we call the resulting number of particles the
\emph{amount of constituent} $i$ and denote it by $n_i$. When a
number-of-particle counting measurement procedure is performed in
a region of space delimited by at least one ideal-surface patch,
some particles may be found across the surface. Therefore, an
outcome of the procedure must also be the sum, for all the
particles in this boundary situation, of a suitably defined
fraction of their spatial extension which is within the given
region of space. As a result, the \emph{number of particles} and
the \emph{amount of constituent i} will not be quantized but will
have continuous spectra.

A level of description $L_2$ is called \emph{deeper} than a level
of description $L_1$ if the amount of every constituent in $L_2$
is conserved for all the physical phenomena considered, whereas
the same is not true for the constituents in $L_1$. For instance,
the atomic level of description is deeper than the chemical one
(because chemical reaction mechanisms do not conserve the number
of molecules of each type, whereas they conserve the number of
nuclei of each type as well as the number of electrons).

 Levels
of description typically have a hierarchical structure whereby the
constituents of a given level are aggregates of the constituents
of a deeper level.

\spazio

\noindent \textbf{Region of space which contains particles of the}
$i$-th \textbf{constituent}. We will call region of space which
contains particles of the $i$-th constituent a connected region
$\textsf{\emph{R}}_i$ of physical space (the three-dimensional
Euclidean space) in which particles of the $i$-th constituent are
contained. The boundary surface of $\textsf{\emph{R}}_i$ may be a
patchwork of \emph{walls}, i.e., surfaces impermeable to particles
of the $i$-th constituent, and ideal surfaces (permeable to
particles of the $i$-th constituent). The geometry of the boundary
surface of $\textsf{\emph{R}}_i$ and its permeability topology
nature (walls, ideal surfaces) can vary in time, as well as the
number of particles contained in $\textsf{\emph{R}}_i$.

\spazio

\noindent \textbf{Collection of matter, composition}. We will call
\emph{collection of matter}, denoted by $\CA$, a set of particles
of one or more constituents which is described by specifying the
allowed reaction mechanisms between different constituents and, at
any time instant $t$, the set of $r$ connected regions of space,
$\RA{} = \textsl{\textsf{R}}^A_1,\dots , \,
\textsl{\textsf{R}}^A_i,\dots, \,\textsl{\textsf{R}}^A_r$, each of
which contains $n^A_i$ particles of a single kind of constituent.
The regions of space $\RA{}$ can vary in time and overlap. Two
regions of space may contain the same kind of constituent provided
that they do not overlap. Thus, the $i$-th constituent could be
identical with the $j$-th constituent, provided that
$\textsl{\textsf{R}}^A_i$ and $\textsl{\textsf{R}}^A_j$ are
disjoint. If, due to time changes, two  regions of space which
contain the same kind of constituent begin to overlap, from that
instant a new collection of matter must be considered.

\spazio

\noindent \emph{Comment}. This method of description allows to
consider the presence of internal walls and/or internal
\emph{semipermeable} membranes, \emph{i.e.}, surfaces which can be
crossed only by some kinds of constituents and not others. In the
simplest case of a collection of matter without internal
partitions, the regions of space $\RA{}$ coincide at every time
instant.

\spazio

 The amount $n_i$ of the
constituent in the $i$-th region of space can vary in time for two
reasons:
\begin{itemize}\item \emph{matter exchange}: during a time interval
in which the boundary surface of $\textsf{\emph{R}}_i$ is not
entirely a wall, particles may be transferred into or out of
$\textsf{\emph{R}}_i$; we denote by
$\dot{\textbf{\emph{n}}}^{A\leftarrow}$ the set of rates at which
particles  are transferred in or out of each region, assumed
positive if inward, negative if outward;
\item \emph{reaction mechanisms}:  in a portion of space where two or more
regions overlap, the allowed reaction mechanisms may transform,
according to well specified proportions (\emph{e.g.},
stoichiometry), particles of one or more regions into particles of
one or more other regions.
\end{itemize}

\spazio

\noindent \textbf{Compatible compositions, set of compatible
compositions}. We say that two compositions,
$\textbf{\emph{n}}^{1A}$ and $\textbf{\emph{n}}^{2A}$ of a given
collection of matter $\CA$ are \emph{compatible} if the change
between $\textbf{\emph{n}}^{1A}$ and $\textbf{\emph{n}}^{2A}$ or
viceversa can take place as a consequence of the allowed reaction
mechanisms without matter exchange. We will call \emph{set of
compatible compositions} for a system $A$ the set of all the
compositions of $A$ which are compatible with a given one. We will
denote a set of compatible compositions for $A$ by the symbol
$(\textbf{\emph{n}}^{0A}$, $\pmb{\nu}^A)$. By this we mean that
the set of $\tau$ allowed reaction mechanisms is defined like for
chemical reactions by a matrix of stoichiometric coefficients
$\pmb{\nu}^A=[\nu_k^{(\ell)}]$, with $\nu_k^{(\ell)}$ representing
the stoichiometric coefficient of the $k$-th constituent in the
$\ell$-th reaction. The set of compatible compositions is a
$\tau$-parameter set defined by the reaction coordinates
$\pmb{\varepsilon}^A=\varepsilon^A_1,\dots,\varepsilon^A_\ell,
\dots,\varepsilon^A_\tau$ through the proportionality relations
\begin{equation}\label{composition} \textbf{\emph{n}}^A =
\textbf{\emph{n}}^{0A} + \pmb{\nu}^A \cdot \pmb{\varepsilon}^A \;
,
\end{equation}
where $\textbf{\emph{n}}^{0A}$ denotes the composition
corresponding to the value zero of all the reaction coordinates
$\pmb{\varepsilon}^A$. To fix ideas and for convenience, we will
select $\pmb{\varepsilon}^A=0$ at time $t=0$ so that
$\textbf{\emph{n}}^{0A}$ is the composition at time $t=0$ and we
may call it the \emph{initial composition}.\\ In general, the rate
of change of the amounts of constituents is  subject to the
\emph{amounts balance equations}
\begin{equation}\label{composition-rate} \dot{\textbf{\emph{n}}}^A =
\dot{\textbf{\emph{n}}}^{A\leftarrow} + \pmb{\nu}^A \cdot
\dot{\pmb{\varepsilon}}^A \; .
\end{equation}

\spazio

\noindent \textbf{External force field}. Let us denote by
\textbf{F} a force field given by the superposition of a
gravitational field \textbf{G}, an electric field \textbf{E}, and
a magnetic induction field \textbf{B}. Let us denote by
$\Sigma_t^A$ the union of all the regions of space $\RA{t}$ in
which the constituents of $\CA$ are contained, at a time instant
$t$, which we also call region of space occupied by $\CA$ at time
$t$. Let us denote by $\Sigma^A$ the union of the regions of space
$\Sigma_t^A$, \emph{i.e.}, the union of all the regions of space
occupied by $\CA$ during its time evolution.

 We call
\emph{external force field} \emph{for} $\CA$ \emph{at time} $t$,
denoted by $\FAt \,$, the spatial distribution of \textbf{F} which
is measured at time $t$ in $\Sigma_t^A$ if all the constituents
and the walls of $\CA$ are removed and placed far away from
$\Sigma_t^A$. We call \emph{external force field} \emph{for}
$\CA$, denoted by $\FA$, the spatial and time distribution of
\textbf{F} which is measured in $\Sigma^A$ if all the constituents
and the walls of $\CA$ are removed and placed far away from
$\Sigma^A$.

\spazio

\noindent \textbf{System, properties of a system}.  We will call
\emph{system} $A$ a collection of matter $\CA$ defined by the
initial composition $\textbf{\emph{n}}^{0A}$,  the stoichiometric
coefficients $\pmb{\nu}^A$ of the allowed reaction mechanisms, and
the possibly time-dependent specification, \emph{over the entire
time interval of interest}, of:
\begin{itemize}\item the geometrical variables and the
nature of the boundary surfaces that define the regions of space
$\RA{t}$, \item the rates
$\dot{\textbf{\emph{n}}}_t^{A\leftarrow}$ at which particles are
transferred in or out of the regions of space, and \item the
external force field distribution $\FAt$ for $\CA$,
\end{itemize} provided that the following conditions apply:
\begin{enumerate}
\item an ensemble of identically prepared replicas of
$\CA$ can be obtained at any instant of time $t$, according to a
specified set of instructions or preparation scheme; \item a set
of measurement procedures, $P^A_1,\dots, P^A_n$, exists, such that
when each $P^A_i$ is applied on replicas of $\CA$ at any given
instant of time $t$: each replica responds with a numerical
outcome which may vary from replica to replica; but either the
time interval $\Delta t$ employed to perform the measurement can
be made arbitrarily short so that the measurement outcomes
considered for $P^A_i$ are those which correspond to the limit as
$\Delta t \rightarrow 0$, or the measurement outcomes are
independent of the time interval $\Delta t$ employed to perform
the measurement;
\item the arithmetic mean $\langle P^A_i\rangle_t$ of the numerical
outcomes of repeated applications of any of these procedures,
$P^A_i$, at an instant $t$, on an ensemble of identically prepared
replicas, is a value which is the same for every subensemble of
replicas of $\CA$ (the latter condition guarantees the so-called
statistical \emph{homogeneity} of the ensemble); $\langle
P^A_i\rangle_t$ is called the \emph{value of} $P^A_i$ for $\CA$ at
time $t$; \item the set of measurement procedures, $P^A_1,\dots,
P^A_n$, is \emph{complete} in the sense that the set of values
$\{\langle P^A_1\rangle_t,\dots, \langle P^A_n\rangle_t\}$ allows
to predict the value of any other measurement procedure satisfying
conditions 2 and 3.
\end{enumerate}
Then, each measurement procedure satisfying conditions 2 and 3 is
called a \emph{property} of system $A$, and the set $P^A_1,\dots ,
P^A_n$ a \emph{complete set of properties} of system $A$.

\spazio

\noindent \emph{Comment}.  Although in general the amounts of
constituents, $\nA{t}$, and the reaction rates,
$\dot{\pmb{\varepsilon}}_t$, are properties according to the above
definition, we will list them separately and explicitly whenever
it is convenient for clarity. In particular, in typical chemical
kinetic models, $\dot{\pmb{\varepsilon}}_t$ is assumed to be a
function of $\nA{t}$ and other properties.

\spazio

\noindent \textbf{State of a system}. Given a system $A$ as just
defined, we call \emph{state of system $A$ at time $t$}, denoted
by $A_t$, the set of the values  \emph{at time $t$} of
\begin{itemize}
\item  all the properties of the system or, equivalently, of
a complete set of properties, $\{\langle P_1\rangle_t,\dots,
\langle P_n\rangle_t\}$,
\item the amounts of constituents, $\nA{t}$, \item the geometrical
variables and the nature of the boundary surfaces of the regions
of space $\RA{t}$,
\item the rates $\dot{\textbf{\emph{n}}}_t^{A\leftarrow}$ of
particle  transfer in or out of the regions of space, and
\item the external force field distribution in the region of space $\Sigma^A_t$
occupied by $A$ at time $t$, $\FAt$. \end{itemize} With respect to
the chosen complete set of properties, we can write
\begin{equation}\label{state}A_t\equiv\left\{\langle P_1\rangle_t,\dots,
\langle P_n\rangle_t;\nA{t};\RA{t};
\dot{\textbf{\emph{n}}}_t^{A\leftarrow}; \FAt \right\} \;\; .
\end{equation} For shorthand, states $A_{t_1}$,
$A_{t_2}$,\dots, are denoted by $A_1$, $A_2$,\dots. Also, when the
context allows it, the value $\langle P^A\rangle_{t_1}$ of
property $P^A$ of system $A$ at time $t_1$ is denoted depending on
convenience by the symbol $P^A_1$, or simply $P_1$.

\spazio

\noindent \textbf{Closed system, open system}. A system $A$ is
called a \emph{closed system} if, at every time instant $t$, the
boundary surface of every region of space
$\textsf{\emph{R}}^A_{it}$ is a wall. Otherwise, $A$ is called an
\emph{open system}.

\spazio

\noindent \emph{Comment}. For a closed system, in each region of
space $\textsf{\emph{R}}^A_i$, the number of particles of the
$i$-th constituent can change only as a consequence of allowed
reaction mechanisms.

\spazio

\noindent \textbf{Composite system, subsystems}. Given a system
$C$ in the external force field $\textbf{F}_e^{C}$, we will say
that $C$ is the \emph{composite} of systems $A$ and $B$, denoted
$AB$, if: (a) there exists a pair of systems $A$ and $B$ such that
the external force field which obtains when both $A$ and $B$ are
removed and placed far away coincides with $\textbf{F}_e^{C}$; (b)
no region of space $\textsl{\textsf{R}}^A_i$ overlaps with any
region of space $\textsl{\textsf{R}}^B_j$; and (c) the
$r_C=r_A+r_B$ regions of space of $C$ are
$\pmb{\textsl{\textsf{R}}}^C = \textsl{\textsf{R}}^A_1,\dots , \,
\textsl{\textsf{R}}^A_i,\dots, \,\textsl{\textsf{R}}^A_{r_A},
\textsl{\textsf{R}}^B_1,\dots , \, \textsl{\textsf{R}}^B_j,\dots,
\,\textsl{\textsf{R}}^B_{r_B} $. Then we say that $A$ and $B$ are
\emph{subsystems} of the \emph{composite system} $C$, and we write
$C=AB$ and denote its state at time $t$ by $C_t=(AB)_t$.

\spazio

\noindent \textbf{Isolated system}. We say that a closed system
$I$ is an isolated system in the stationary external force field
$\FI$, or simply an \emph{isolated system}, if, during the whole
time evolution of $I$: (a) only the particles of $I$ are present
in $\Sigma^I$; (b) the external force field for $I$, $\FI$, is
stationary, \emph{i.e.}, time independent and conservative.

\spazio

\noindent\emph{Comment.} In simpler words, a system $I$ is
isolated if, at every time instant: no other material particle is
present in the whole region of space $\Sigma^I$ which will be
crossed by system $I$ during its time evolution;  if system $I$ is
removed, only a stationary (vanishing or non-vanishing)
conservative force field is present in $\Sigma^I$.

\spazio

\noindent \textbf{Separable closed systems}. Consider a composite
system $AB$, with $A$ and $B$ closed subsystems. We say that
systems $A$ and $B$ are \emph{separable} at time $t$ if:
\begin{itemize}
\item the force field external to $A$ coincides (where defined)
with the force field external to $AB$, i.e., $\FAt = \FABt$; \item
the force field external to $B$ coincides (where defined) with the
force field external to $AB$, i.e., $\FBt = \FABt$.
\end{itemize}

\spazio

\noindent\emph{Comment.} In simpler words, system $A$ is separable
from  $B$ at time $t$, if at that instant the force field produced
by $B$ is vanishing in the region of space occupied by $A$ and
viceversa.  During the subsequent time evolution of $AB$, $A$ and
$B$ need not remain separable at all times.

\spazio

\noindent \textbf{Subsystems in uncorrelated states}. Consider a
composite system $AB$ such that at time $t$ the states $A_t$ and
$B_t$ of the two subsystems fully determine the state $(AB)_t$,
\emph{i.e.}, the values of all the properties of $AB$ can be
determined by \emph{local} measurements of properties of systems
$A$ and $B$. Then, at time $t$, we say that the states of
subsystems $A$ and $B$ are \emph{uncorrelated from each other}, and we write the
state of $AB$ as $(AB)_t=A_t B_t$. We also say, for brevity, that
$A$ and $B$ are \emph{systems uncorrelated from each other} at time $t$.

\spazio

\noindent \textbf{Correlated states, correlation}.  If at time $t$
the states $A_t$ and $B_t$ do not fully determine the state
$(AB)_t$ of the composite system $AB$, we say that $A_t$ and $B_t$
are \emph{states correlated with each other}. We also say, for
brevity, that $A$ and $B$ are \emph{systems correlated with each
other} at time $t$.

\spazio

\noindent\emph{Comment.} Two systems $A$ and $B$ which are
uncorrelated from each other at time $t_1$ can undergo an
interaction such that they are correlated with each other at time
$t_2 > t_1$.

\spazio

\noindent \emph{Comment. Correlations between isolated systems.}
Let us consider an isolated system $I = AB$ such that, at time
$t$, system $A$ is separable and uncorrelated from $B$. This
circumstance does not exclude that, at time $t$, $A$ and/or $B$
(or both) may be correlated with a system $C$, even if the latter
is isolated, \emph{e.g.} it is far away from the region of space
occupied by $AB$. Indeed our definitions of separability and
correlation are general enough to be fully compatible with the
notion of quantum correlations, \emph{i.e.}, \emph{entanglement},
which  plays an important role in modern physics. In other words,
assume that an isolated system $U$ is made of three subsystems
$A$, $B$, and $C$, \emph{i.e.}, $U=ABC$, with $C$ isolated and
$AB$ isolated. The fact that $A$ is uncorrelated from $B$, so that
according to our notation we may write $(AB)_t=A_t B_t$, does not
exclude that $A$ and $C$ may be entangled, in such a way that the
states $A_t$ and $C_t$ do not determine the state of $AC$,
\emph{i.e.}, $(AC)_t\ne A_t C_t$, \emph{nor} we can write
$U_t=(A)_t (BC)_t$.

\spazio

\noindent \textbf{Environment of a system, scenario}. If for the
time span of interest a system $A$ is a subsystem of an isolated
system $I=AB$, we can choose $AB$ as the isolated system to be
studied. Then, we will call $B$ the \emph{environment} of $A$, and
we call $AB$ the \emph{scenario} under which $A$ is studied.

\spazio

\noindent \emph{Comment}. The chosen scenario $AB$  contains as
subsystems all and only the systems that are allowed to interact
with $A$; thus all the remaining systems in the universe, even if
correlated with $AB$, are considered as not available for
interaction.

\spazio

\noindent \emph{Comment. A system uncorrelated from its
environment in one scenario, may be correlated with its
environment in a broader scenario.} Consider a system $A$ which,
in the scenario $AB$, is uncorrelated from its environment $B$ at
time $t$. If at time $t$ system $A$ is entangled with an isolated
system $C$, in the scenario $ABC$, $A$ is correlated with its
environment $BC$.

\spazio

\noindent \textbf{Process, cycle.} We  call \emph{process} for a
system $A$  from state $A_1$ to state $A_2$ in the scenario $AB$,
denoted by $(AB)_1 \rightarrow (AB)_2$, the change of state from
$(AB)_1$ to $(AB)_2$ of the isolated system $AB$ which defines the
scenario. We call \emph{cycle} for a system $A$  a process whereby
the final state $A_2$ coincides with the initial state $A_1$.

\spazio

\noindent \emph{Comment}. In every process of any system $A$, the
force field $\FAB$ external to $AB$, where $B$ is the environment
of $A$, cannot change. In fact, $AB$ is an isolated system and, as
a consequence, the force field external to $AB$ is stationary.
Thus, in particular, for all the states in which a  system $A$ is
separable:
\begin{itemize}
\item the force field $\FAB$ external to $AB$, where $B$ is the
environment of $A$, is the same; \item the force field $\FA$
external to $A$ coincides, where defined, with the force field
$\FAB$ external to $AB$, \emph{i.e.}, the force field produced by
$B$ (if any) has no effect on $A$.
\end{itemize}

\spazio

\noindent \textbf{Process between uncorrelated states, external
effects.} A process in the scenario $AB$ in which the end states
of system $A$ are both uncorrelated from its environment $B$ is
called \emph{process between uncorrelated states} and denoted by
$\Pi_{12}^{A,B}\equiv (A_1 \rightarrow A_2)_{B_1 \rightarrow
B_2}$. In such a process,  the change of state of the environment
$B$ from $B_1$ to $B_2$ is called \emph{effect external to} $A$.
Traditional expositions of thermodynamics consider only this kind
of process.

\spazio

\noindent \textbf{Composite process}. A time-ordered sequence of
processes between uncorrelated states of a system $A$ with
environment $B$, $\Pi_{1k}^{A,B}=$ ($\Pi_{12}^{A,B}$,
$\Pi_{23}^{A,B}$,\dots, $\Pi_{(i-1)i}^{A,B}$,\dots,
$\Pi_{(k-1)k}^{A,B}$) is called a \emph{composite process}  if the
final state of $AB$ for process $\Pi_{(i-1)i}^{A,B}$ is the
initial state of $AB$ for process $\Pi_{i(i+1)}^{A,B}$, for $i =
1, 2, \dots , k-1$. When the context allows the simplified
notation $ \Pi_{i}$ for $i = 1, 2, \dots , k-1$ for the processes
in the  sequence, the \emph{composite process} may also be denoted
by ($\Pi_{1}$, $\Pi_{2}$,\dots , $\Pi_{i}$,\dots, $\Pi_{k-1}$).

\spazio

\noindent \textbf{Reversible process, reverse of a reversible
process}. A process for $A$ in the scenario $AB$, $(AB)_1
\rightarrow (AB)_2$, is called  a \emph{reversible process} if
there exists a process $(AB)_2 \rightarrow (AB)_1$ which restores
the initial state of the isolated system $AB$. The process $(AB)_2
\rightarrow (AB)_1$ is called \emph{reverse} of process $(AB)_1
\rightarrow (AB)_2$. With different words, a process of an
isolated system $I = AB$ is reversible if it can be reproduced as
a part of a cycle of the isolated system $I$.
 For a reversible
process between uncorrelated states, $\Pi_{12}^{A,B}\equiv (A_1
\rightarrow A_2)_{B_1 \rightarrow B_2}$, the  \emph{reverse} will
be denoted by
 $-\Pi^{A,B}_{12}\equiv (A_2 \rightarrow A_1)_{B_2 \rightarrow B_1}$.

\spazio

\noindent \textit{Comment}. The reverse process may be achieved in
more than one way (in particular, not necessarily by retracing the
sequence of states $(AB)_t$, with $t_1\le t\le t_2$, followed by
the isolated system $AB$ during the forward process).

\spazio

\noindent \textit{Comment. The reversibility in one scenario does
not grant the reversibility in another.} If the smallest isolated
system which contains $A$ is $AB$ and another isolated system $C$
exists in a different region of space, one can choose as
environment of $A$ either $B$ or $BC$. Thus, the time evolution of
$A$ can be described by the process $(AB)_1 \rightarrow (AB)_2$ in
the scenario $AB$ or by the process $(ABC)_1 \rightarrow (ABC)_2$
in the scenario $ABC$. For instance, the process $(AB)_1
\rightarrow (AB)_2$ could be irreversible, however by broadening
the scenario so that interactions between $AB$ and $C$ become
available,  a reverse process $(ABC)_2 \rightarrow (ABC)_1$ may be
possible. On the other hand, a process $(ABC)_1 \rightarrow
(ABC)_2$ could be irreversible on account of an irreversible
evolution $C_1\rightarrow C_2$ of $C$, even if the process $(AB)_1
\rightarrow (AB)_2$ is reversible.

\spazio

\noindent \emph{Comment. A reversible process need not be slow}.
In the general framework we are setting up, it is noteworthy that
nowhere we state nor we need the  concept that a process to be
reversible needs to be \emph{slow} in some sense. Actually, as
well represented in \citep{GB2005} and clearly understood within
dynamical systems models based on linear or nonlinear master
equations, the time evolution of the state of a system is the
result of a competition between  (hamiltonian) mechanisms which
are reversible and  (dissipative) mechanisms which are not. So, to
design a reversible process in the nonequilibrium domain, we most
likely need a \emph{fast} process, whereby the state is changed
quickly by a fast hamiltonian dynamics, leaving negligible time
for the dissipative mechanisms to produce  irreversible effects.

\spazio

\noindent \textbf{Weight}. We call  \emph{weight} a system $M$ always
separable and uncorrelated from its environment, such that:
\begin{itemize}
\item $M$ is closed, it has a single constituent contained in a
single region of space whose shape and volume are fixed, \item  it
has a constant  mass $m$;
\item in any process, the difference between the initial and the
final state of $M$ is determined uniquely by the change in the
position $z$ of the center of mass of $M$, which can move only
along a straight line whose direction is identified by the unit
vector $\textbf{k}=\nabla z$;
\item along
the straight line there is a uniform stationary external
gravitational force field $\textbf{G}_e=-g\textbf{k}$, where $g$ is
a constant gravitational acceleration.
\end{itemize}
As a consequence, the difference in potential energy between any
initial and final states of $M$ is given by $mg(z_2-z_1)$.

\spazio

\noindent \textbf{Weight process, work in a weight process}. A
process between states of a closed system $A$ in which $A$
is separable and uncorrelated from its environment
is called a \emph{weight process}, denoted by $(A_1 \rightarrow
A_2)_W$, if the only effect external to  $A$ is the displacement
of the center of mass of a weight $M$ between two  positions $z_1$
and $z_2$.  We call \emph{work performed by $A$ (or, done by A) in the weight
process}, denoted by the symbol $W_{12}^{A\rightarrow}$, the
quantity
\begin{equation}
W_{12}^{A\rightarrow} = mg(z_2-z_1) \;\; .
\end{equation}
Clearly, the \emph{work done by} $A$ is positive if $z_2>z_1$ and
negative if $z_2<z_1$. Two equivalent symbols for the opposite of
this work, called \emph{work received by} $A$, are
$-W_{12}^{A\rightarrow} = W_{12}^{A\leftarrow}$.

\spazio

\noindent \textbf{Equilibrium state  of a closed system}. A state
$A_t$ of a closed system $A$, with environment $B$,  is called an
\emph{equilibrium state} if:\begin{itemize} \item $A$ is a
separable system at time $t$; \item state $A_t$ does not change
with time;  \item  state $A_t$ can be reproduced while $A$ is an
isolated system in the external force field $\FA$, which
coincides, where defined, with $\FAB$. \end{itemize}

\spazio

\noindent \textbf{Stable equilibrium state of a closed system}. An
equilibrium state of a closed system $A$ in which $A$ is
uncorrelated from its environment $B$, is called a \emph{stable
equilibrium state} if it cannot be modified by any process between
states in which $A$ is separable and uncorrelated from its
environment such that neither  the geometrical configuration of
the walls which bound the regions of space $\RA{}$ where the
constituents of $A$ are contained, nor the state of the
environment $B$ of $A$ have net changes.

\spazio

\noindent \textit{Comment. The stability of equilibrium in one
scenario does not grant the stability of equilibrium in another.}
Consider a system $A$ which, in the scenario $AB$, is uncorrelated
from its environment $B$ at time $t$ and is in a stable
equilibrium state. If at time $t$ system $A$ is entangled with an
isolated system $C$, then in the scenario $ABC$, $A$ is correlated
with its environment $BC$, therefore, our definition of stable
equilibrium state is not satisfied.

\section{\label{first}Definition of energy for a closed system}

\noindent \textbf{First Law}. Every pair of states ($A_1$, $A_2$)
of a closed system $A$ in which $A$ is separable and uncorrelated
from its environment can be interconnected by means of a weight
process for $A$. The works performed by the system in any two
weight processes between the same initial and final states are
identical.

\spazio

\noindent \textbf{Definition of energy for a closed system. Proof
that it is a property}. Let ($A_1$, $A_2$) be any pair of states
of a closed system $A$ in which $A$ is separable and uncorrelated from
its environment. We call
\emph{energy difference} between states $A_2$ and $A_1$ either the
work $W_{12}^{A\leftarrow}$ received by $A$ in any weight process
from $A_1$ to $A_2$ or the work $W_{21}^{A\rightarrow}$ done by
$A$ in any weight process from $A_2$ to $A_1$; in symbols:
\begin{equation}\label{energy}
E^A_2 - E^A_1 = W_{12}^{A\leftarrow} \quad \mbox{or}\quad E^A_2 -
E^A_1 = W_{21}^{A\rightarrow} .
\end{equation}
The first law guarantees that at least one of the weight processes
considered in Eq. (\ref{energy}) exists. Moreover, it yields the
following consequences: \\ (\emph{a}) if both weight processes
$(A_1 \rightarrow A_2)_W$ and $(A_2 \rightarrow A_1)_W$ exist, the
two forms of
    Eq. (\ref{energy}) yield the same result
    ($W_{12}^{A\leftarrow}=W_{21}^{A\rightarrow}$);\\
(\emph{b}) the energy difference between two states $A_2$ and $A_1$
in which $A$ is separable and uncorrelated from its environment depends only on the states $A_1$ and
$A_2$;\\
(\emph{c}) (\emph{additivity of energy differences for separable
 systems uncorrelated from each other}) consider a pair of closed systems $A$ and
$B$; if $A_1B_1$ and $A_2B_2$ are states of the composite system $AB$
such that $AB$ is separable and uncorrelated from its environment and, in addition, $A$ and $B$ are separable and uncorrelated from each other, then
\begin{equation}\label{additivity}
E^{AB}_2 - E^{AB}_1 = E^A_2 - E^A_1 + E^B_2 - E^B_1 \;\; ;
\end{equation}
(\emph{d}) (\emph{energy is a property for every separable
system uncorrelated from its environment}) let $A_0$ be a reference state of a closed system $A$ in which $A$ is separable and uncorrelated from its environment, to which we assign
an arbitrarily chosen value of energy $E^A_0$; the value of the energy of $A$ in any other
state $A_1$ in which $A$ is separable  and uncorrelated from its environment
is determined uniquely by the equation
\begin{equation}\label{energyabs}
E^A_1 = E^A_0 + W_{01}^{A\leftarrow} \quad \mbox{or}\quad E^A_1 =
E^A_0 + W_{10}^{A\rightarrow}
\end{equation}
where $W_{01}^{A\leftarrow}$ or $W_{10}^{A\rightarrow}$ is the
work in any  weight process for $A$ either from $A_0$ to $A_1$ or
from $A_1$ to $A_0$; therefore, energy is a property of $A$.\\
Rigorous proofs of these consequences can be found in
\citep{GB2005,Zanchini:1986}, and will not be repeated here. In
the proof of Eq. (\ref{additivity}), the restrictive condition of
the absence of correlations between $AB$ and its environment as
well as between $A$ and $B$, implicit in \citep{GB2005} and
\citep{Zanchini:1986}, can be released by means of an assumption
(Assumption 3) which is presented and discussed in the next
section. As a result, Eq. (\ref{additivity}) holds also if
$(AB)_1$ e $(AB)_2$ are arbitrarily chosen states of the composite
system $AB$, provided that $AB$, $A$ and $B$ are separable
systems.

\section{\label{second}Definition of thermodynamic entropy
 for a closed system}

\noindent \textbf{Assumption 1: restriction to normal system}. We
will call \emph{normal system} any system $A$ that, starting from
every state in which it is separable and uncorrelated from its
environment, can be changed to a non-equilibrium state with higher
energy by means of a weight process for $A$ in which the regions
of space $\RA{}$ occupied by the constituents of $A$ have no net
change (and $A$ is again separable and uncorrelated from its
environment).

From here on, we consider only normal systems; even when we say
only \emph{system} we mean a \emph{normal system}.

\spazio

\noindent \emph{Comment}. For a normal system, the energy is
unbounded from above; the system can accommodate an indefinite
amount of energy, such as when its constituents have
translational, rotational or vibrational degrees of freedom. In
traditional treatments of thermodynamics, Assumption 1 is
\emph{not stated explicitly, but it is used}, for example when one
states that any amount of work can be transferred to a thermal
reservoir by a stirrer. Notable exceptions to this assumption are
important quantum theoretical model systems, such as spins,
qubits, qudits, etc. whose energy is bounded from above. The
extension of our treatment to such so-called \emph{special
systems} is straightforward, but we omit it here for simplicity.

\spazio

\noindent \textbf{Theorem 1. Impossibility of a PMM2}. If  a
normal system $A$ is in a stable equilibrium state, it is
impossible to lower its energy by means of a weight process for
$A$ in which the regions of space $\RA{}$ occupied by the
constituents of $A$ have no net change.

\noindent \textbf{Proof}. Suppose that, starting from a stable
equilibrium state $A_{se}$ of $A$, by means of a weight process
$\Pi_1$ with positive work $W^{A\rightarrow}=W>0$, the energy of
$A$ is lowered and the regions of space $\RA{}$ occupied by the
constituents of $A$ have no net change. On account of Assumption
1, it would be possible to perform a weight process $\Pi_2$ for
$A$ in which the regions of space $\RA{}$ occupied by the
constituents of $A$ have no net change, the weight $M$ is restored
to its initial state so that the positive amount of energy
$W^{A\leftarrow}=W>0$ is supplied back to $A$, and the final state
of $A$ is a nonequilibrium state, namely, a state clearly
different from $A_{se}$. Thus, the zero-work composite process
($\Pi_1$, $\Pi_2$) would violate the definition of stable
equilibrium state.

\spazio

\noindent \emph{Comment. Kelvin-Planck statement of the Second
Law}. As  noted in  \citep{6} and \citep[p.64]{GB2005}, the
impossibility of a perpetual motion machine of the second kind
(PMM2), which is also known as the \emph{Kelvin-Planck statement
of the Second Law}, is a corollary of the definition of stable
equilibrium state, provided that we adopt the (usually implicitly)
restriction to normal systems (Assumption 1).

\spazio

\noindent \textbf{Second Law}. Among all the states in which a
closed system $A$ is separable and uncorrelated from its
environment and the constituents of $A$ are contained in a given
set of regions of space $\RA{}$, there is a stable equilibrium
state for every value of the energy $E^A$.

\spazio

\noindent \textbf{Lemma 1. Uniqueness of the stable equilibrium
state}. There can be no pair of
different stable equilibrium states of a closed system $A$ with
identical regions of space $\RA{}$ and the same value of the
energy $E^A$.

\noindent \textbf{Proof}. Since $A$ is closed and in any stable
equilibrium state it is separable and uncorrelated from its
environment, if two such states existed, by the first law and the
definition of energy they could be interconnected by means of a
zero-work weight process. So, at least one of them could be
changed to a different state with no external effect, and hence
would not satisfy the definition of stable equilibrium state.

\spazio

\noindent \emph{Comment}. Recall that for a closed system, the
composition $\nA{}$ belongs to the  set of compatible compositions
$(\textbf{\emph{n}}^{0A}$, $\pmb{\nu}^A)$ fixed once and for all
by the definition of the system.

\spazio

\noindent \emph{Comment. Statements of the Second Law}. The
combination of our statement of the Second Law and Lemma 1
establishes, for a closed system whose matter is constrained into
given regions of space, the existence and uniqueness of a stable
equilibrium state for every value of the energy; this proposition
is known as the \emph{Hatsopoulos-Keenan statement of the Second
Law} \citep{6}. Well-known historical statements of the Second
Law, in addition to the Kelvin-Planck statement discussed above,
are due to Clausius and to Carath\'{e}odory. In  \citep[p.64,
p.121, p.133]{GB2005} it is  shown that each of  these historical
statements is a logical consequence of the Hatsopoulos-Keenan
statement combined with a further assumption, essentially
equivalent to our Assumption 2 below.

\spazio

\noindent \textbf{Lemma 2}. Any stable equilibrium state $A_s$ of
a closed system $A$ is accessible via an irreversible zero-work
weight process from any other state $A_1$ in which
$A$ is separable and uncorrelated with its environment and has the
same regions of space $\RA{}$ and the same value of the energy
$E^A$.

\noindent \textbf{Proof}. By the first law and the definition of
energy, $A_s$ and $A_1$ can be interconnected by a zero-work
weight process for $A$. However, a zero-work weight process from
$A_s$ to $A_1$ would violate the definition of stable equilibrium
state. Therefore, the process must be in the direction from $A_1$
to $A_s$. The absence of a zero-work weight process in the
opposite direction, implies that any zero-work weight process from
$A_1$ to $A_s$ is irreversible.

\spazio

\noindent \textbf{Corollary 1}. Any state in which a closed system
$A$ is separable and uncorrelated from its environment can be
changed to a unique stable equilibrium state by means of a
zero-work weight process for $A$ in which the regions of space
$\RA{}$ have no net change.

\noindent \textbf{Proof.} The thesis follows immediately from the
Second Law, Lemma 1 and Lemma 2.\\

\noindent \textbf{Mutual stable equilibrium states}. We say that
two stable equilibrium states $A_{\rm se}$ and $B_{\rm se}$ are
\emph{mutual stable equilibrium states} if, when $A$ is in state
$A_{\rm se}$ and $B$ in state $B_{\rm se}$, the composite system
$AB$ is in a stable equilibrium state. The definition holds also
for a pair of states of the same system: in this case, system $AB$
is composed of $A$ and of a duplicate of $A$.

\spazio

\noindent \textbf{Identical copy of a system}. We say that a
system $A^d$, always separable from $A$ and uncorrelated with $A$,
is an \emph{identical copy} of system $A$ (or, a \emph{duplicate}
of $A$) if, at every time instant:
\begin{itemize}
\item the difference between the set of regions of
space $\RA{}^d$ occupied by the matter of $A^d$ and that $\RA{}$
occupied by the matter of $A$ is only a rigid translation $\Delta
\textbf{r}$ with respect to the reference frame considered, and
the composition of $A^d$ is compatible with that of $A$; \item the
external force field for $A^d$ at any position $\textbf{r} +
\Delta \textbf{r}$ coincides with the external force field for $A$
at the position $\textbf{r}$.
\end{itemize}

\spazio

\noindent \textbf{Thermal reservoir}. We call \emph{thermal
reservoir} a system $R$ with a single constituent, contained in a
fixed region of space, with a vanishing external force field, with
energy values restricted to a finite range such that in any of its
stable equilibrium states, $R$ is in mutual stable equilibrium
with an identical copy of $R$, $R^d$, in any of its stable
equilibrium states.

\spazio

\noindent \emph{Comment}. Every single-constituent system without
internal boundaries and applied external fields, and with a number
of particles of the order of one mole (so that the \emph{simple
system} approximation as defined in  \citep[p.263]{GB2005}
applies), when restricted to a fixed region of space of
appropriate volume and to the range of energy values corresponding
to the so-called \emph{triple-point} stable equilibrium states, is
an excellent approximation of a thermal reservoir.

\spazio

\noindent \textbf{Reference thermal reservoir}. A thermal
reservoir chosen once and for all, will be called a
\emph{reference thermal reservoir}. To fix ideas, we will choose
as our reference thermal reservoir one having water as
constituent, with a volume, an amount, and a range of energy
values which correspond to the so-called \emph{solid-liquid-vapor
triple-point} stable equilibrium states.

\spazio

\noindent \textbf{Standard weight process}. Given a pair of states
$(A_1, A_2)$ of a closed system $A$, in which $A$ is separable and
uncorrelated from its environment, and a thermal reservoir $R$, we
call \emph{standard weight process} for $AR$ from $A_1$ to $A_2$ a
weight process for the composite system $AR$ in which the end
states of $R$ are stable equilibrium states. We denote by
$(A_1R_1\rightarrow A_2R_2)^{\sw}$ a standard weight process for
$AR$ from $A_1$ to $A_2$ and by $\DE{R}{A}$ the corresponding
energy change of the thermal reservoir $R$.

\spazio

\noindent \textbf{Assumption 2}. Every pair of states ($A_1$,
$A_2$) in which a closed system $A$ is separable and uncorrelated
from its environment can be interconnected by a reversible
standard weight process for $AR$, where $R$ is an arbitrarily
chosen thermal reservoir.

\spazio

\noindent \textbf{Theorem 2}. For a given closed system $A$ and a
given  reservoir $R$, among all the standard weight processes for
$AR$ between a given pair of states ($A_1$, $A_2$) in which system
$A$ is separable and uncorrelated from its environment, the energy
change $\DE{R}{A}$ of the thermal reservoir $R$ has a lower bound
which is reached if and only if the process is reversible.

\noindent \textbf{Proof}. Let $\Pi_{AR}$ denote  a standard weight
process for $AR$ from $A_1$ to $A_2$,  and $\Pi_{AR\rm rev}$ a
reversible one;  the energy changes of $R$ in processes $\Pi_{AR}$
and $\Pi_{AR\rm rev}$ are, respectively,  $\DE{R}{A}$ and
$\DErev{R}{A}$. With the help of Figure 1, we will prove that,
regardless of the initial state of $R$:\\ a) $\DErev{R}{A} \leq
\DE{R}{A}$;
\\ b) if also $\Pi_{AR}$ is reversible, then $\DErev{R}{A} =
\DE{R}{A}$;\\ c)  if $\DErev{R}{A} = \DE{R}{A}$, then also
$\Pi_{AR}$ is reversible.

\noindent \textbf{Proof of a)}. Let us denote by $R_1$ and by
$R_2$ the initial and the final states of $R$ in process
$\Pi_{AR\rm rev}$. Let us denote by $R^d$ the duplicate of $R$
which is employed in process $\Pi_{AR}$, by $R^d_3$ and by $R^d_4$
the initial and the final states of $R^d$ in this process. Let us
suppose, \emph{ab absurdo}, that  $\DErev{R}{A} >  \DE{R}{A}$.
Then, the composite process ($-\Pi_{AR\rm rev}$, $\Pi_{AR}$) would
be a weight process for $RR^d$ in which, starting from the stable
equilibrium state $R_2R^d_3$, the energy of $RR^d$ is lowered and
the regions of space occupied by the constituents of $RR^d$ have
no net change, in contrast with Theorem 1. Therefore,
$\DErev{R}{A} \leq \DE{R}{A}$.

\noindent \textbf{Proof of b)}. If $\Pi_{AR}$ is reversible too,
then, in addition to $\DErev{R}{A} \leq \DE{R}{A}$, the relation
$\DE{R}{A} \leq \DErev{R}{A}$ must hold too. Otherwise, the
composite process ($\Pi_{AR\rm rev}$,  $-\Pi_{AR}$) would be a
weight process for $RR^d$ in which, starting from the stable
equilibrium state $R_1R^d_4$, the energy of $RR^d$ is lowered and
the regions of space occupied by the constituents of $RR^d$ have
no net change, in contrast with Theorem 1. Therefore,
$\DErev{R}{A} = \DE{R}{A}$.

\noindent \textbf{Proof of c)}. Let $\Pi_{AR}$ be a standard
weight process for $AR$, from $A_1$ to $A_2$, such that $\DE{R}{A}
= \DErev{R}{A}$, and let $R_1$ be the initial state of $R$ in this
process. Let $\Pi_{AR\rm rev}$ be a reversible standard weight
process for $AR$, from $A_1$ to $A_2$, with the same initial state
$R_1$ of $R$. Thus,  $R^d_3$ coincides with $R_1$ and $R^d_4$
coincides with $R_2$. The composite process ($\Pi_{AR}$,
$-\Pi_{AR\rm rev}$) is a cycle for the isolated system $ARB$,
where $B$ is the environment of $AR$. As a consequence, $\Pi_{AR}$
is reversible, because it is a part of a cycle of the isolated
system $ARB$.

\spazio

\begin{figure}
\includegraphics[width=0.35\textwidth]{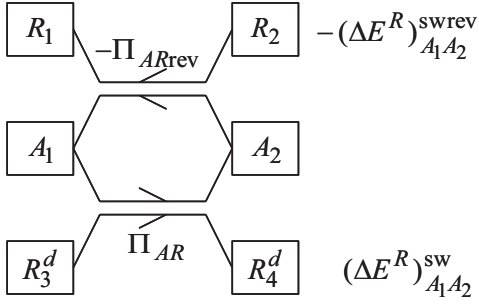}
\label{Figure1} \caption{Illustration of the proof of Theorem 2:
standard weight processes  $\Pi_{AR\rm rev}$ (reversible) and
$\Pi_{AR}$; $R^d$ is a duplicate of $R$; see text.}\end{figure}
\begin{figure}\includegraphics[width=0.35\textwidth]{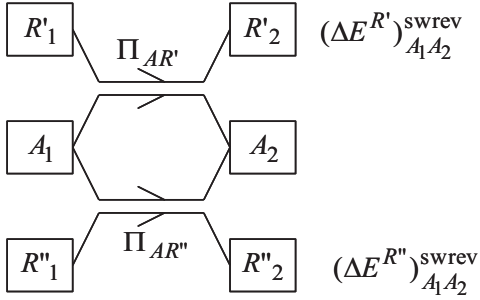}
\label{Figure2} \caption{Illustration of the proof of Theorem 3,
part a): reversible standard weight processes  $\Pi_{AR'}$ and
$\Pi_{AR''}$, see text.}
\end{figure}


\noindent \textbf{Theorem 3}. Let $R'$ and $R''$ be any two
thermal reservoirs and consider the energy changes,
$\DErev{R'}{A}$ and $\DErev{R''}{A}$ respectively, in the
reversible standard  weight processes $\Pi_{AR'}=(A_1R'_{1
}\rightarrow A_2R'_{2 })^{\rm swrev}$ and $\Pi_{AR''}=(A_1R''_{1
}\rightarrow A_2R''_{2 })^{\rm swrev}$, where ($A_1$, $A_2$) is an
arbitrarily chosen pair of states of any closed system $A$ in
which $A$ is separable and uncorrelated from  its environment.
Then the ratio $\DErev{R'}{A}/\DErev{R''}{A}$\,:\\ a) is
positive;\\ b) depends only on $R'$ and $R''$, \emph{i.e.}, it is
independent of \emph{(i)} the initial stable equilibrium states of
$R'$ and $R''$, \emph{(ii)} the choice of system $A$, and
\emph{(iii)} the choice of states $A_1$ and $A_2$.

\noindent \textbf{Proof of a)}. With the help of Figure 2, let us
suppose that $\DErev{R'}{A}  < 0$. Then, $\DErev{R''}{A}$ cannot
be zero. In fact, in that case the composite process ($\Pi_{AR'}$,
$-\Pi_{AR''}$), which is a cycle for $A$, would be a weight
process for $R'$ in which, starting from the stable equilibrium
state $R'_1$, the energy of $R'$ is lowered and the region of
space occupied by $R'$ has no net change, in contrast with Theorem
1. Moreover, $\DErev{R''}{A}$ cannot be positive. In fact, if it
were positive, the work performed by $R'R''$ as a result of the
overall weight process ($\Pi_{AR'}$, $-\Pi_{AR''}$) for $R'R''$
would be
\begin{equation}\label{work-R'R''}
W^{R'R''\rightarrow} = - \DErev{R'}{A} + \DErev{R''}{A} \;\; ,
\end{equation}
where both terms are positive. On account of Assumption 1 and
Corollary 1, after the process ($\Pi_{AR'}$,  $-\Pi_{AR''}$), one
could perform a weight process $\Pi_{R''}$ for $R''$ in which a
positive amount of energy  equal to $\DErev{R''}{A}$ is given back
to $R''$ and the latter is restored to its initial stable
equilibrium state. As a result, the composite process  ($\Pi_{AR'}$,
$-\Pi_{AR''}$, $\Pi_{R''}$) would be a weight process for $R'$ in
which, starting from the stable equilibrium state $R'_1$, the
energy of $R'$ is lowered and the regions of space occupied by the
constituents of $R'$ have no net change, in contrast with Theorem
1. Therefore, the assumption $\DErev{R'}{A} < 0$ implies
$\DErev{R''}{A}  < 0$.\\ Let us suppose that $\DErev{R'}{A} > 0$.
Then, for process $-\Pi_{AR'}$  one has $\DErevinv{R'}{A} < 0$. By
repeating the previous argument, one proves that for process
$-\Pi_{AR''}$  one has $\DErevinv{R''}{A} < 0$. Therefore, for
process  $\Pi_{AR''}$  one has $\DErev{R''}{A} > 0$.

\begin{figure*}\label{Figure3}
\includegraphics[width=0.8\textwidth]{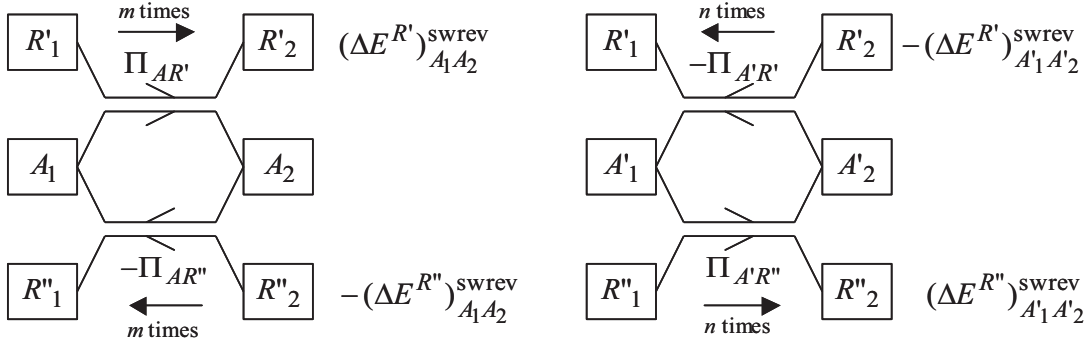}
\caption{Illustration of the proof of Theorem 3, part b):
composite processes $\Pi_A$ and $\Pi_{A'}$, see text.}
\end{figure*}

\noindent \textbf{Proof of b)}. Given a pair of states ($A_1$,
$A_2$) of a closed system $A$, consider  the reversible standard
weight process $\Pi_{AR'}=(A_1R'_{1 }\rightarrow A_2R'_{2 })^{\rm
swrev}$ for $AR'$, with $R'$ initially in state $R'_1$, and the
reversible standard weight process $\Pi_{AR''}=(A_1R''_{1
}\rightarrow A_2R''_{2 })^{\rm swrev}$ for $AR''$, with $R''$
initially in state $R''_1$. Moreover, given a pair of states
($A'_1$, $A'_2$) of another closed system $A'$, consider the
reversible standard weight process $\Pi_{A'R'}=(A'_1R'_{1
}\rightarrow A'_2R'_{2 })^{\rm swrev}$ for $A'R'$, with $R'$
initially in state $R'_1$, and the reversible standard weight
process $\Pi_{A'R''}=(A'_1R''_{1 }\rightarrow A'_2R''_{2 })^{\rm
swrev}$ for $A'R''$, with $R''$ initially in state $R''_1$.

With the help of Figure 3, we will prove that the changes in
energy of the reservoirs in these processes obey the relation
\begin{equation}\label{energyratios}
\frac{\DErev{R'}{A}}{\DErev{R''}{A}} =
\frac{\DErev{R'}{A'}}{\DErev{R''}{A'}}\;\; .
\end{equation}

Let us assume:  $\DErev{R'}{A}>0$ and $\DErev{R'}{A'} > 0$, which
implies,  $\DErev{R''}{A}>0$ and $\DErev{R''}{A'} > 0$ on account
of part a) of the proof. This is not a restriction, because it is
possible to reverse the processes under exam. Now, as is well
known, any real number can be approximated with an arbitrarily
high accuracy by a rational number. Therefore, we will assume that
the energy changes $\DErev{R'}{A}$ and $\DErev{R'}{A'} $ are
rational numbers, so that whatever is the value of their ratio,
there exist two positive integers $m$ and $n$ such that
$\DErev{R'}{A}/\DErev{R'}{A'} = n/m$, \emph{i.e.},
\begin{equation}\label{firstreservoir}
m \; \DErev{R'}{A} = n \; \DErev{R'}{A'} \;\; .
\end{equation}
Therefore, as sketched in Figure 3, let us consider the composite
processes $\Pi_A$ and $\Pi_A'$ defined as follows.  $\Pi_A$ is the
following composite weight process for system $A R' R''$: starting
from the initial state $R'_1$ of $R'$ and $R''_2$ of $R''$, system
$A$ is brought from $A_1$ to $A_2$ by a reversible standard weight
process for $AR'$, then from $A_2$ to $A_1$ by a reversible
standard weight process for $AR''$; whatever the new states of
$R'$ and $R''$ are, again  system $A$ is brought from $A_1$ to
$A_2$ by a reversible standard weight process for $AR'$ and back
to $A_1$ by a reversible standard weight process for $AR''$, until
the cycle for $A$ is repeated $m$ times. Similarly, $\Pi_{A'}$ is
a composite weight processes for system $A' R' R''$ whereby
starting from the end states of $R'$ and $R''$ reached by $\Pi_A$,
system $A'$ is brought from $A'_1$ to $A'_2$ by a reversible
standard weight process for $A'R''$, then from $A'_2$ to $A'_1$ by
a reversible standard weight process for $A'R'$; and so on until
the cycle for $A'$ is repeated $n$ times.\\ Clearly, the whole
composite process ($\Pi_A$, $\Pi_A\,'$) is a cycle for $AA'$.
Moreover, it is a cycle also for $R'$. In fact, on account of
Theorem 2, the energy change of $R'$ in each process $\Pi_{AR'}$
is equal to $\DErev{R'}{A}$ regardless of its initial state, and
in each process $-\Pi_{A'R'}$ the energy change of $R'$ is equal
to $-\DErev{R'}{A'}$. Therefore, the energy change of $R'$ in the
composite process ($\Pi_A$, $\Pi_A'$) is $m \; \DErev{R'}{A}  - n
\; \DErev{R'}{A'}$ and equals zero on account of Eq.
(\ref{firstreservoir}). As a result, after ($\Pi_A$, $\Pi_A'$),
reservoir $R'$ has been restored to its initial state, so that
($\Pi_A$, $\Pi_A'$) is a reversible weight process for $R''$.\\
Again on account of Theorem 2, the overall energy change of $R''$
in ($\Pi_A$, $\Pi_A'$) is $- m \; \DErev{R''}{A} + n \;
\DErev{R''}{A}$. If this quantity were negative, Theorem 1 would
be violated. If this quantity were positive, Theorem 1 would also
be violated by the reverse of the process, ($-\Pi_A'$,  $-\Pi_A$).
Therefore, the only possibility is that $- m \; \DErev{R''}{A} + n
\; \DErev{R''}{A} = 0$, \emph{i.e.},
\begin{equation}\label{secondreservoir}
m \; \DErev{R''}{A} = n \; \DErev{R''}{A'} \;\; .
\end{equation}
Finally,  taking the ratio of Eqs. (\ref{firstreservoir}) and
(\ref{secondreservoir}), we obtain Eq. (\ref{energyratios}) which
is our conclusion.

\spazio

\noindent \textbf{Temperature of a thermal reservoir}. Let $R$ be
a given thermal reservoir and $R^o$ a reference thermal reservoir.
Select an arbitrary pair of states ($A_1$, $A_2$) in which an
arbitrary closed system $A$ is separable and uncorrelated from its
environment, and consider the energy changes $\DErev{R}{A}$ and
$\DErev{R^o}{A}$ in two reversible standard weight processes from
$A_1$ to $A_2$, one for $AR$ and the other for $AR^o$,
respectively. We call \emph{temperature} of $R$ the positive
quantity
\begin{equation}\label{temperature}
T_R = T_{R^o} \; \frac{\DErev{R}{A}}{\DErev{R^o}{A}} \;\; ,
\end{equation}
where $T_{R^o}$ is a positive constant associated arbitrarily with
the reference thermal reservoir $R^o$.  If for $R^o$ we select a
thermal reservoir having water as constituent, with energy
restricted to the solid-liquid-vapor triple-point range, and we
set $T_{R^o}=273.16$ K, we obtain the unit kelvin (K) for the
thermodynamic temperature, which is adopted in the International
System of Units (SI). Clearly, the temperature $T_R$ of $R$ is
defined only up to an arbitrary multiplicative constant.

\spazio

\noindent \textbf{Corollary 2}. The ratio of the temperatures of
two thermal reservoirs, $R'$ and $R''$, is independent of the
choice of the reference thermal reservoir and can be measured
directly as
\begin{equation}\label{temperatureratio}
\frac{T_{R'}}{T_{R''}} = \frac{\DErev{R'}{A}}{\DErev{R''}{A}} \;\;
,
\end{equation}
where $\DErev{R'}{A}$ and $\DErev{R''}{A}$ are the energy changes
of $R'$ and $R''$ in two reversible standard weight processes, one
for $AR'$ and the other for $AR''$, which interconnect the same
but otherwise arbitrary pair of states ($A_1$, $A_2$) in which a
closed system $A$ is separable and uncorrelated from its
environment.

\noindent \textbf{Proof}. Let $\DErev{R^o}{A}$ be the energy
change of the reference thermal reservoir $R^o$ in any reversible
standard weight process for $AR^o$ which interconnects the same
states ($A_1$, $A_2$) of $A$. From Eq. (\ref{temperature}) we have
\begin{equation}\label{TRfirst}
T_{R\,'} = T_{R^o} \; \frac{\DErev{R'}{A}}{\DErev{R^o}{A}} \;\; ,
\end{equation}
\begin{equation}\label{TRsecond}
T_{R\,''} = T_{R^o} \; \frac{\DErev{R''}{A}}{\DErev{R^o}{A}} \;\;
,
\end{equation}
therefore the ratio of Eqs. (\ref{TRfirst}) and (\ref{TRsecond})
yields Eq. (\ref{temperatureratio}).

\spazio

\noindent \textbf{Corollary 3}. Let ($A_1$, $A_2$) be any pair of
states in which a closed system $A$ is separable and uncorrelated
from its environment, and let $\DErev{R}{A}$ be the energy change
of a thermal reservoir $R$ with temperature $T_R$, in any
reversible standard weight process for $AR$ from $A_1$ to $A_2$.
Then, for the given system $A$, the ratio $\DErev{R}{A} / T_R$
depends only on the pair of states ($A_1$, $A_2$), \emph{i.e.}, it
is independent of the choice of reservoir $R$ and of its initial
stable equilibrium state $R_{1 }$.

\noindent \textbf{Proof}. Let us consider two  reversible standard
weight processes from $A_1$ to $A_2$, one for  $AR'$ and the other
for $AR''$, where $R'$ is a thermal reservoir with temperature
$T_{R'}$ and $R''$ is a thermal reservoir with temperature
$T_{R''}$. Then, equation (\ref{temperatureratio}) yields
\begin{equation}\label{energy-temperature}
\frac{\DErev{R'}{A}}{T_{R'}} = \frac{\DErev{R''}{A}}{T_{R''}} \;\;
.
\end{equation}

\spazio

\noindent \textbf{Definition of (thermodynamic) entropy for a
closed system. Proof that it is a property}. Let ($A_1$ , $A_2$)
be any pair of  states in which a closed system $A$ is separable
and uncorrelated from its environment $B$, and let $R$ be an
arbitrarily chosen thermal reservoir placed in $B$. We call
\emph{entropy difference} between $A_2$ and $A_1$ the quantity
\begin{equation}\label{entropy}
S^A_2 - S^A_1 = - \frac{\DErev{R}{A}}{T_R}
\end{equation}
where $\DErev{R}{A}$ is the energy change of $R$ in any reversible
standard weight process for $AR$ from $A_1$ to $A_2$, and $T_R$ is
the temperature of $R$. On account of Corollary 3, the right hand
side of Eq. (\ref{entropy}) is determined uniquely by states $A_1$
and $A_2$.\\
Let $A_0$ be a reference  state in which $A$ is separable and
uncorrelated from its environment, to which we assign an
arbitrarily chosen value of entropy $S^A_0$. Then, the value of
the entropy of $A$ in any other  state $A_1$ in which $A$ is
separable and uncorrelated from its environment, is determined
uniquely by the equation
\begin{equation}\label{entropyabs}
S^A_1 = S^A_0  - \frac{\DErevzero{R}{A}}{T_R} \;\; ,
\end{equation}
where $\DErevzero{R}{A}$ is the energy change of $R$ in any
reversible standard weight process for $AR$ from $A_0$ to $A_1$,
and $T_R$ is the temperature of $R$. Such a process exists for
every state $A_1$, on account of Assumption 2. Therefore, entropy
is a property of $A$ and is defined for every state of $A$ in
which $A$ is separable and uncorrelated from its environment.

\spazio

\noindent \textbf{Theorem 4. Additivity of entropy differences for
uncorrelated states}. Consider the pairs of states $(C_1 = A_1B_1,
C_2 = A_2 B_2)$ in which the composite system $C =A B$ is
separable and uncorrelated from its environment, and  systems $A$
and $B$ are separable and uncorrelated from each other. Then,
\begin{equation}\label{entropyadditivity}
S^{AB}_{A_2 B_2} - S^{AB}_{A_1 B_1} = S^A_2 - S^A_1 + S^B_2 -
S^B_1 \;\; .
\end{equation}

\noindent \textbf{Proof}. Let us choose a thermal reservoir $R$,
with temperature $T_R$, and consider the composite process ($\Pi_{AR}$,
$\Pi_{BR}$)  where $\Pi_{AR}$ is a reversible standard weight
process for $AR$ from $A_1$ to $A_2$, while $\Pi_{BR}$ is a
reversible standard weight process for $BR$ from $B_1$ to $B_2$.
The composite process ($\Pi_{AR}$, $\Pi_{BR}$) is a reversible standard
weight process for $CR$ from $C_1$ to $C_2$, in which the energy
change of $R$ is the sum of the energy changes in the constituent
processes $\Pi_{AR}$ and $\Pi_{BR}$, \emph{i.e.},
$\DErev{R}{C}=\DErev{R}{A}+\DErev{R}{B}$. Therefore:
\begin{equation}\label{entropyadditivity-proof}
\frac{\DErev{R}{C}}{T_R} = \frac{\DErev{R}{A}}{T_R} +
\frac{\DErev{R}{B}}{T_R} \;\; .
\end{equation}
Equation (\ref{entropyadditivity-proof}) and the definition of
entropy (\ref{entropy}) yield Eq. (\ref{entropyadditivity}).

\spazio

\noindent \emph{Comment}. As a consequence of Theorem 4, if the
values of entropy are chosen so that they are additive in the
reference states, entropy results as an additive property. Note,
however, that the proof of additivity requires that $(A_1, B_1)$
and $(A_2, B_2)$ are pairs of states such that the subsystems $A$
and $B$ are uncorrelated from each other.

\spazio

\noindent \textbf{Theorem 5}. Let ($A_1$, $A_2$) be any pair of
states in which a closed system $A$ is separable and uncorrelated
from its environment and let $R$ be a
thermal reservoir with temperature $T_R$. Let $\Pi_{AR\rm irr}$ be
any irreversible standard weight process for $AR$ from $A_1$ to
$A_2$ and let $\DEirr{R}{A}$ be the energy change of $R$ in this
process. Then
\begin{equation}\label{inequality}
 - \frac{\DEirr{R}{A}}{T_R} < S^A_2 - S^A_1 \;\; .
\end{equation}

\noindent \textbf{Proof}. Let $\Pi_{AR\rm rev}$ be any reversible
standard weight process for $AR$ from $A_1$ to $A_2$ and let
$\DErev{R}{A}$ be the energy change of $R$ in this process. On
account of Theorem 2,
\begin{equation}\label{minimum-energychange}
\DErev{R}{A} < \DEirr{R}{A} \;\; .
\end{equation}
Since $T_R$ is positive, from Eqs. (\ref{minimum-energychange})
and (\ref{entropy}) one obtains
\begin{equation}\label{inequality-two}
 - \frac{\DEirr{R}{A}}{T_R} <  - \frac{\DErev{R}{A}}{T_R} = S^A_2 - S^A_1\;\; .
\end{equation}

\spazio

\noindent \textbf{Theorem 6. Principle of entropy nondecrease}.
Let $(A_1, A_2)$ be a pair of  states in which a closed system $A$
is separable and uncorrelated from its environment and let $(A_1 \rightarrow A_2)_W$ be any weight
process for $A$ from $A_1$ to $A_2$. Then, the entropy difference
$S^A_2 - S^A_1$ is equal to zero if and only if the weight process
is reversible; it is strictly positive if and only if the weight
process is irreversible.

\noindent \textbf{Proof}. If  $(A_1 \rightarrow A_2)_W$ is
reversible, then it is a special case of a reversible standard
weight process for $AR$ in which the initial stable equilibrium
state of $R$ does not change. Therefore, $\DErev{R}{A}=0$ and by
applying the definition of entropy, Eq. (\ref{entropy}), one
obtains
\begin{equation}\label{nondecrease-rev}
S^A_2 - S^A_1 = - \frac{\DErev{R}{A}}{T_R} = 0 \;\; .
\end{equation}
If  $(A_1 \rightarrow A_2)_W$ is irreversible,
then it is a special case of an irreversible standard weight
process for $AR$ in which the initial stable equilibrium state of
$R$ does not change. Therefore, $\DEirr{R}{A}=0$ and Equation
(\ref{inequality})  yields
\begin{equation}\label{nondecrease-irr}
S^A_2 - S^A_1 > - \frac{\DEirr{R}{A}}{T_R} = 0 \;\; .
\end{equation}
Moreover: if a weight process $(A_1 \rightarrow A_2)_W$ for $A$ is
such that $S^A_2 - S^A_1 = 0$, then the process must be
reversible, because we just proved that for any irreversible
weight process  $S^A_2 - S^A_1 > 0$; if a weight process $(A_1
\rightarrow A_2)_W$ for $A$ is such that $S^A_2 - S^A_1 > 0$, then
the process must be irreversible, because we just proved that for
any reversible weight process $S^A_2 - S^A_1 = 0$.

\spazio

\noindent \textbf{Corollary 4}. If states $A_1$ and $A_2$ can be
interconnected by means of a reversible weight process for $A$,
they have the same entropy. If states $A_1$ and $A_2$ can be
interconnected by means of a zero-work reversible weight process
for $A$, they have the same energy and the same entropy.

\noindent \textbf{Proof}. These are straightforward  consequences
of  Theorem 6 together with the definition of energy.

\spazio

\noindent \textbf{Theorem 7. Highest-entropy principle}. Among all
the states of a closed system $A$ such that $A$ is separable and
uncorrelated from its environment, the constituents of $A$ are
contained in a given set of regions of space $\RA{}$ and the value
of the energy $E^A$ of $A$ is fixed, the entropy of $A$ has the
highest value only in the unique stable equilibrium state $A_{se}$
determined by $\RA{}$ and $E^A$.

 \noindent \textbf{Proof}. Let
$A_g$ be any other state of $A$ in the set of states considered
here. On account of the first law and of the definition of energy,
$A_g$ and $A_{se}$ can be interconnected by a zero work weight
process for $A$, either $(A_g \rightarrow A_{se})_W$ or $(A_{se}
\rightarrow A_g)_W$. However, the existence of a zero work weight
process $(A_{se} \rightarrow A_g)_W$ would violate the definition
of stable equilibrium state. Therefore, a zero work weight process
$(A_g \rightarrow A_{se})_W$ exists and is irreversible, so that
Theorem 6 implies $S^A_{se} > S^A_g $.

\spazio

\noindent \textbf{Assumption 3. Existence of spontaneous
decorrelations and impossibility of spontaneous creation of
correlations}. Consider a system $AB$ composed of two closed
subsystems $A$ and $B$. Let $(AB)_1$ be a state in which $AB$ is
separable and uncorrelated from its environment and such that in
the corresponding states $A_1$ and $B_1$, systems $A$ and $B$ are
separable but correlated; let $A_1B_1$ be the state of $AB$ such
that the corresponding states $A_1$ and $B_1$ of $A$ and $B$ are
the same as for state $(AB)_1$, but $A$ and $B$ are uncorrelated.
Then, a zero work weight process $((AB)_1 \rightarrow  A_1B_1)_W$
for $AB$ is possible, while a weight process $( A_1B_1 \rightarrow
(AB)_1)_W$ for $AB$ is impossible.

\spazio


\noindent \textbf{Corollary 5. Energy difference between states of
a composite system in which subsystems are correlated with each
other}. Let $(AB)_1$ and $(AB)_2$ be states of a composite system
$AB$ in which $AB$ is separable and uncorrelated from its
environment, while systems $A$ and $B$ are separable but
correlated with each other. We have
\begin{eqnarray}\label{energydifference-correlated}
E^{AB}_{(AB)_2} - E^{AB}_{(AB)_1} &=& E^{AB}_{A_2B_2} -
E^{AB}_{A_1B_1}\nonumber \\&=& E^A_2 - E^A_1 + E^B_2 - E^B_1 \;\;
.
\end{eqnarray}

\noindent \textbf{Proof}. Since a zero work weight process
$((AB)_1 \rightarrow A_1B_1)_W$ for $AB$ exists on account of
Assumption 3, states $(AB)_1$ and $A_1B_1$ have the same energy.
In other words, the energy of a composite system in state $(AB)_1$
with separable but correlated subsystems coincides with the energy
of the composite system in state $A_1B_1$ where its separable
subsystems are uncorrelated in the corresponding states $A_1$ and
$A_2$.

\spazio

\noindent \textbf{Definition of energy for a state in which a
system is correlated with its environment}. On account of Eq.
(\ref{energydifference-correlated}), we will say that the energy
of a system $A$ in a state $A_1$ in which $A$ is correlated with
its environment is equal to the energy of system $A$ in the
corresponding state $A_1$ in which $A$ is uncorrelated from its
environment.

\spazio

\noindent \emph{Comment}. Equation
(\ref{energydifference-correlated}) and the definition of energy
for a state in which a system is correlated with its environment
extend the definition of energy and the proof of the additivity of
energy differences presented in  \citep{GB2005,Zanchini:1986} to
the case in which systems $A$ and $B$ are separable but correlated
with each other.\\ To our knowledge, Assumption 3 (never made
explicit) underlies all reasonable models of relaxation and
decoherence.

\spazio

\noindent \textbf{Corollary 6. De-correlation entropy}. Given a
pair of (different) states $(AB)_1$ and $A_1B_1$ as defined in
Assumption 3, then we have
 \begin{equation}\label{correlationS}
\sigma^{AB}_{(AB)_1} = S^{AB}_{A_1B_1}- S^{AB}_{(AB)_1} > 0 \;\; ,
\end{equation}
where the positive quantity $\sigma^{AB}_1$ is called the
\emph{de-correlation entropy}\footnote{Explicit expressions of
this property in the quantum formalism are given, \emph{e.g.}, in
\citet{wehrl,QTb,demon}.} of state $(AB)_1$. Clearly, if the
subsystems are uncorrelated, \emph{i.e.}, if $(AB)_1=A_1B_1$, then
$\sigma^{AB}_{(AB)_1} = \sigma^{AB}_{A_1B_1} = 0$.

 \noindent \textbf{Proof}. On account of Assumption 3, a zero work
weight process $\Pi_{AB} = ((AB)_1 \rightarrow  A_1B_1)_W$ for
$AB$ exists. Process $\Pi_{AB}$ is irreversible, because the
reversibility of $\Pi_{AB}$ would require the existence of a zero
work weight process for $AB$ from $A_1B_1$ to $(AB)_1$, which is
excluded by Assumption 3. Since $\Pi_{AB}$ is irreversible,
Theorem 6 yields the conclusion.

\spazio

\noindent \emph{Comment.} Let $(AB)_1$ and $(AB)_2$ be a pair of
states of a composite system $AB$ such that $AB$ is separable and
uncorrelated from its environment, while subsystems $A$ and $B$
are separable but correlated with each other. Let $A_1B_1$ and
$A_2B_2$ be the corresponding pairs of states of $AB$, in which
the subsystems $A$ and $B$ are in the same states as before, but
are uncorrelated from each other. Then, the entropy difference
between $(AB)_2$ and $(AB)_1$ is not equal to the entropy
difference between $A_2B_2$ and $A_1B_1$ and therefore, on account
of Eq. (\ref{entropyadditivity}), it is not equal to the sum of
the entropy difference between $A_2$ and $A_1$ and the entropy
difference between $B_2$ and $B_1$, evaluated in the corresponding
states in which subsystems $A$ and $B$ are uncorrelated from each
other. In fact, combining Eq. (\ref{entropyadditivity}) with Eq.
(\ref{correlationS}), we have
 \begin{eqnarray}\label{additivitycorrelationS}
S^{AB}_{(AB)_2} - S^{AB}_{(AB)_1}
&=&(S^A_2-S^A_1)+(S^B_2-S^B_1)\nonumber \\ && -
  ( \sigma^{AB}_{(AB)_2} - \sigma^{AB}_{(AB)_1}) \;\; .
\end{eqnarray}

\section{\label{fundamental}Fundamental relation, temperature, and
Gibbs relation for closed systems}

\noindent \textbf{Set of equivalent stable equilibrium states}.
We will call \emph{set of equivalent stable
equilibrium states} of a closed system $A$, denoted $ESE^A$, a
subset of its stable equilibrium states such that any pair of
states in the set:
\begin{itemize}
\item  differ from one another by some geometrical features of the regions of
space $\RA{}$; \item have the same composition; \item can be
interconnected by a zero-work reversible weight process for $A$
and, hence, by Corollary 4, have the same  energy and the same
entropy.
   \end{itemize}

\spazio

\noindent \emph{Comment}. Let us recall that, for all the stable
equilibrium states of a closed system $A$ in a scenario $AB$,
system $A$ is separable and the external force field $\FA$ =
$\FAB$ is the same; moreover, all the compositions of $A$ belong
to the same set of compatible compositions
$(\textbf{\emph{n}}^{0A}, \nu^A)$.

\spazio

\noindent \textbf{Parameters of a closed system}. We will call
\emph{parameters} of a closed system $A$, denoted by
$\pmb{\beta}^A=\beta^A_1, \dots , \beta^A_s$, a minimal set of
real variables sufficient to fully and uniquely parametrize all
the different sets of equivalent stable equilibrium states $ESE^A$
of $A$. In the following, we will consider systems with a finite
number $s$ of parameters.

\spazio

 \noindent \emph{Examples}. Consider a system
$A$ consisting of a single particle confined in spherical region
of space of volume $V$; the box is centered at position
$\textsf{r}$ which can move in a larger region where there are no
external fields. Then, it is clear that any rotation or
translation of the spherical box within the larger region can be
effected in a zero-work weight process that does not alter the
rest of the state. Therefore, the position of the
center of the box is \emph{not} a parameter of the system. The
volume instead is a parameter. The same holds if the box is cubic.
If it is a parallelepiped, instead, the parameters are the sides
$\ell_1$, $\ell_2$, $\ell_3$ but not its position and orientation.
For a more complex geometry of the box, the parameters are any
minimal set of geometrical features sufficient to fully describe
its shape, regardless of its position and orientation. The same if
instead of one, the  box contains many particles.\\ Suppose now we
have a spherical box, with one or many particles, that can be
moved in a larger region where there are $k$ subregions, each much
larger than the box and each with an external  electric field
everywhere parallel to the $x$ axis and with uniform magnitude
$E_{ek}$. As part of the definition of the system, let us restrict
it only to the states such that the box is fully contained in one
of these regions. For this system, the magnitude of $E_e$ can be
changed in a weight process by moving $A$ from one uniform field
subregion to another, but this in general will vary the energy.
Therefore, in addition to the volume of the sphere, this system
will have $k$ as a parameter identifying the subregion where the
box is located. Equivalently, the subregion can be identified by
the parameter $E_e$ taking values in the set $\{E_{ek}\}$. For
each value of the energy $E$, system $A$ has a set $ESE^A$ for
every pair of values of the parameters ($V$, $E_e$) with  $E_e$ in
$\{E_{ek}\}$.

\spazio

\noindent \textbf{Corollary 7. Fundamental relation for the stable
equilibrium states of a closed system}. On the set of all the
stable equilibrium states of a closed system $A$ (in scenario
$AB$, for given initial composition $\textbf{\emph{n}}^{0A}$,
stoichiometric coefficients $\pmb{\nu^A}$ and external force field
$ \FA$), the entropy is given by a single valued function
\begin{equation}\label{fundamental-relation}
S^A_{\rm se} = S^A_{\rm se}(E^A, \pmb{\beta}^A) \;\; ,
\end{equation}
which is called \emph{fundamental relation} for the stable
equilibrium states of $A$. Moreover, also the reaction coordinates
are given by a single valued function
\begin{equation}\label{chem-eq-composition}
\pmb{\varepsilon}^A_{\rm se} = \pmb{\varepsilon}^A_{\rm se}(E^A,
\pmb{\beta}^A) \;\; ,
\end{equation}
which specifies the unique composition compatible with the initial
composition $\textbf{\emph{n}}^{0A}$, called the \emph{chemical
equilibrium composition}.

\noindent \textbf{Proof}. On account of the Second Law and Lemma
1, among all the  states of a closed system $A$ with energy $E^A$,
the regions of space $\RA{}$ identify a unique stable equilibrium
state. This implies the existence of a single valued function
$A_{\rm se} = A_{\rm se}(E^A,\RA{})$, where $A_{\rm se}$ denotes
the state, in the sense of Eq. (\ref{state}). By definition, for
each value of the energy $E^A$, the values of the parameters
$\pmb{\beta}^A$ fully identify all the regions of space $\RA{}$
that correspond to a set of equivalent stable equilibrium states
$ESE^A$, which have the same value of the entropy and the same
composition. Therefore, the values of $E^A$ and $\pmb{\beta}^A$
fix uniquely the values of $S^A_{\rm se}$ and of
$\pmb{\varepsilon}^A_{\rm se}$. This implies the existence of the
single valued functions written in Eqs.
(\ref{fundamental-relation}) and (\ref{chem-eq-composition}).

\spazio

\noindent \emph{Comment}.  Clearly, for a non-reactive closed
system, the composition is fixed and equal to the initial,
\emph{i.e.}, $\pmb{\varepsilon}^A_{\rm se}(E^A,
\pmb{\beta}^A)=0$.\\ Usually \citep{6,GB2005}, in view of the
equivalence that defines them, each set $ESE^A$ is thought of as a
single state called \quot{a stable equilibrium state} of $A$.
Thus, for a given closed system $A$ (and, hence,  given initial
amounts of constituents), it is commonly stated that the energy
and the parameters of $A$ determine \quot{a unique stable
equilibrium state} of $A$, which is called \quot{the  chemical
equilibrium state} of $A$ if the system is reactive according to a
given set of stoichiometric coefficients. For a discussion of the
implications of Eq. (\ref{chem-eq-composition}) and its reduction
to more familiar chemical equilibrium criteria in terms of
chemical potentials see, \emph{e.g.}, \citep{BG:04}.

\spazio

\noindent \textbf{Assumption 4}. The fundamental relation
(\ref{fundamental-relation}) is continuous and differentiable with
respect to each of the variables $E^A$ and $\pmb{\beta}^A$.

\spazio

\noindent \textbf{Theorem 8}. For any closed system, for fixed
values of the parameters the fundamental relation
(\ref{fundamental-relation}) is a strictly increasing function of
the energy.

\noindent \textbf{Proof}.  Consider two stable equilibrium states
$A_{{\rm se}1}$ and $A_{{\rm se}2}$ of a closed system $A$, with
energies $E^A_1$ and $E^A_2$, entropies $S^A_{{\rm se}1}$ and
$S^A_{{\rm se}2}$, and with the same regions of space occupied by
the constituents of $A$ (and therefore the same values of the
parameters). Assume $E^A_2>E^A_1$. By Assumption 1, we can start
from state $A_{{\rm se}1}$ and, by a weight process for $A$ in
which the regions of space occupied by the constituents of $A$
have no net changes, add work so that the system ends in a
non-equilibrium state $A_2$ with energy $E^A_2$. By Theorem 6, we
must have $S^A_2\ge S^A_{{\rm se}1}$. Now, on account of Lemma 2,
we can go from state $A_2$ to $A_{{\rm se}2}$ with a zero-work
irreversible weight process for $A$. By Theorem 6, we must have
$S^A_{{\rm se}2}> S^A_2$. Combining the two inequalities, we find
that $E^A_2>E^A_1$ implies $S^A_{{\rm
se}2}> S^A_{{\rm se}1}$.

\spazio

\noindent \textbf{Corollary 8}. The fundamental relation for any
closed system $A$ can be rewritten in the form
\begin{equation}\label{fundamental-relation-two}
E^A_{\rm se} = E^A_{\rm se}(S^A, \pmb{\beta}^A)
 \;\; .
\end{equation}

\noindent \textbf{Proof}. By Theorem 8, for fixed $\pmb{\beta}^A$,
Eq. (\ref{fundamental-relation}) is a strictly increasing function
of $E^A$. Therefore, it is invertible with respect to $E^A$ and,
as a consequence, can be written in the form
(\ref{fundamental-relation-two}).

\spazio

\noindent \textbf{Temperature of a closed system in a stable equilibrium
state}. Consider a stable equilibrium state $A_{\rm se}$ of a
closed system $A$ identified by the  values of $E^A$ and
$\pmb{\beta}^A$. The  partial derivative of the fundamental
relation (\ref{fundamental-relation-two}) with respect to $S^A$,
is denoted by
\begin{equation}\label{temperature-system}
T^A = \bigg (\frac{\partial E^A_{\rm se}}{\partial S^A} \bigg
)_{\pmb{\beta}^A} \;\; .
\end{equation}
Such derivative is always defined on account of Assumption 45.
When evaluated at the  values of $E^A$ and $\pmb{\beta}^A$ that
identify state $A_{\rm se}$, it yields a value that we call the
\emph{temperature} of state $A_{\rm se}$.

\spazio

\noindent \emph{Comment}. One can prove \citep[p.127]{GB2005} that
two stable equilibrium states $A_1$ and $A_2$ of a closed
 system $A$ are mutual stable equilibrium states if and only
if they have the same temperature, \emph{i.e.}, if $T^A_1 =
T^A_2$. Moreover, it is easily proved \citep[p.136]{GB2005} that,
when applied to a thermal reservoir $R$, Eq.
(\ref{temperature-system}) yields that all the stable equilibrium
states of a thermal reservoir have the same temperature which is
equal to the temperature $T_R$ of $R$ defined by Eq.
(\ref{temperature}).

\spazio

\noindent \textbf{Corollary 9}. For any stable equilibrium state
of any (normal) closed system, the temperature is non-negative.

\noindent \textbf{Proof}. The thesis follows immediately from the
definition of temperature, Eq. (\ref{temperature-system}), and
Theorem 8.

\spazio

\noindent \textbf{Gibbs equation for a non-reactive closed
system}. By differentiating Eq. (\ref{fundamental-relation-two}),
one obtains (omitting the superscript \quot{$A$} and the subscript
\quot{se} for simplicity)
\begin{equation}\label{Gibbs}
dE = T\, dS + \sum_{j\,=1}^s F_j \, d\beta_j \;\; ,
\end{equation}
where $F_j$ is called generalized force conjugated to the $j$-th
parameter of $A$, $F_j= \big (\partial E_{\rm se}/\partial \beta_j
\big)_{S,\pmb{\beta}'}$. If all the regions of space $\RA{}$
coincide and the volume $V$ of any of them is a parameter, the
negative of the conjugated generalized force is called
\emph{pressure}, denoted by $p$, $p=- \big (\partial E_{\rm
se}/\partial V
\big)_{S,\pmb{\beta}'}$.

\spazio

\noindent \textbf{Fundamental relation in the quantum formalism}.
Let us recall that the measurement procedures that define energy
and entropy must be applied, in general, to a (homogeneous)
ensemble of identically prepared replicas of the system of
interest. Because the numerical outcomes may vary (fluctuate) from
replica to replica, the values of the energy and the entropy
defined by these procedures are arithmetic means. Therefore, what
we have denoted so far, for simplicity, by the symbols $E^A$ and
$S^A$ should be understood as $\langle E^A\rangle$ and $\langle
S^A\rangle$. Where appropriate, like in the quantum formalism
implementation, this more precise notation should be preferred.
Then, written in full notation, the fundamental relation
(\ref{fundamental-relation}) for a closed system is
\begin{equation}\label{fundamental-relation-full}
\langle S^A\rangle_{\rm se} = S^A_{\rm se}(\langle E^A\rangle,
\pmb{\beta}^A) \;\; ,
\end{equation}
and the corresponding Gibbs relation
\begin{equation}\label{Gibbs-full}
d\langle E\rangle = T\, d\langle S\rangle + \sum_{j\,=1}^s  F_j
 \, d\beta_j \;\; .
\end{equation}

\section{\label{open}Definitions of energy and entropy for an open system}

Our definition of energy is based on the First Law, by which a
weight process is possible between any pair of states $A_1$ and
$A_2$ in which a closed system $A$ is separable and uncorrelated
from its environment. Our definition of entropy is based on
Assumption 2, by which a reversible standard weight process for
$AR$ is possible between any pair of states $A_1$ and $A_2$ in
which a closed system $A$ is separable and uncorrelated from its
environment. In both cases, $A_1$ and $A_2$ have compatible
compositions. In this section, we extend the definitions of energy
and entropy to a set of states in which an open system $O$ is
separable and uncorrelated from its environment; two such
states of $O$ have, in general, non-compatible compositions.

\spazio

\noindent \textbf{Separable open system uncorrelated from its
environment}. Consider an open system $O$ that has $Q$ as its
(open) environment, \emph{i.e.}, the composite system $OQ$ is
isolated in $\FOQ$. We say that system $O$ is \emph{separable}
from $Q$ at time $t$ if the state $(OQ)_t$ of $OQ$ can be
reproduced as (\emph{i.e.}, coincides with)  a state $(AB)_t$ of
an isolated system $AB$ in $\FAB=\FOQ$ such that $A$ and $B$ are
closed and separable at time $t$. If the state $(AB)_t=A_tB_t$,
\emph{i.e.}, is such that $A$ and $B$ are uncorrelated from each
other, then we say that the open system $O$ is \emph{uncorrelated
from its environment} at time $t$, and we have $O_t=A_t$,
$Q_t=B_t$, and $(OQ)_t=O_tQ_t$.

\spazio

\noindent \textbf{Set of elemental species}.  Following
\citep[p.545]{GB2005}, we will call \emph{set of elemental
species} a \emph{complete} set of \emph{independent} constituents
with the following features: (1) (\emph{completeness})  there
exist reaction mechanisms by which all other constituents can be
formed starting only from constituents in the set; and (2)
(\emph{independence}) there exist no reaction mechanisms that
involve only constituents in the set.\\ For example, in chemical
thermodynamics we form a set of elemental species by selecting
among all the chemical species formed by atomic nuclei of a single
kind those that have the most stable molecular structure and form
of aggregation at standard temperature and pressure.

\spazio

\noindent \textbf{Energy and entropy of a separable open system
uncorrelated from its environment}. Let $OQ$ be an isolated system
in $\FOQ$, with $O$ and $Q$ open systems, and let us choose
scenario $OQ$, so that $Q$ is the environment of $O$. Let us
suppose that $O$ has $r$ single-constituent regions of space and a
set of allowed reaction mechanisms with stoichiometric
coefficients $\pmb{\nu}^O$. Let us consider a state $O_1$ in which
$O$ is separable and uncorrelated from its environment and has
composition $\nO{1} = (n^O_1, \dots , n^O_i, \dots , n^O_r)_1$.
Let $A^{\nO{1}}B$ be an isolated system in
$\textbf{F}_e^{A^{\nO{1}}B} = \FOQ$, such that $A^{\nO{1}}$ is
closed, has the same allowed reaction mechanisms as $O$ and
compositions compatible with $\nO{1}$. Let $A^{\nO{1}}_1$ be a
state of $A^{\nO{1}}$ such that, in that state, system
$A^{\nO{1}}$ is a separable system in $\FA^{\nO{1}} =
\textbf{F}_e^{A^{\nO{1}}B}$ and is uncorrelated from its
environment; moreover, the state $A^{\nO{1}}_1$ coincides with
$O_1$,\emph{ i.e.}, has the same values of all the properties. We
will define as energy and entropy of $O$, in state $O_1$, the
energy and the entropy of $A^{\nO{1}}$ in state $A^{\nO{1}}_1$,
namely $E^O_1 = E^{A^{\nO{1}}}_1$ and $S^O_1=S^{A^{\nO{1}}}_1$.
The existence of system $A^{\nO{1}}$ and of state $A^{\nO{1}}_1$
is granted by the
definition of separability for $O$ in state $O_1$.

The values of the energy and of the entropy of $A^{\nO{1}}$, in
state $A^{\nO{1}}_1$, are determined by choosing a reference state
$A^{\nO{1}}_0$ of $A^{\nO{1}}$ and by applying Eqs.
(\ref{energyabs}) and (\ref{entropyabs}). The reference state
$A^{\nO{1}}_0$ and the reference values $E_0^{A^{\nO{1}}}$ and
$S_0^{A^{\nO{1}}}$ are selected as defined
below.

We choose $A^{\nO{1}}$ as the composite of $q$ closed subsystems,
$A^{\nO{1}} = A^1A^2 \cdots A^i \cdots  A^q$, each one containing
an elemental species, chosen so that the composition of
$A^{\nO{1}}$ is compatible with that of $O$ in state $O_1$. Each
subsystem, $A^i$, contains $n_i$ particles of the $i$-th elemental
species and is constrained by a wall in a spherical box with a
variable volume $V^{A^i}$; each box is very far from the others
and is placed in a position where the external force field
$\FA^{\nO{1}}$ is vanishing.

We choose the reference state $A^{\nO{1}}_0$ to be such that  each
subsystem $A^i$ is in a stable equilibrium state $A^i_0$ with a
prescribed temperature, $T_0$, and a volume $V^{A^i}_0$ such that
the pressure has a prescribed value $p_0$.

 We fix the reference values of
the energy and the entropy of the reference state $A^{\nO{1}}_0$
as follows:
\begin{equation}\label{reference-energy}
E_0^{A^{\nO{1}}} = \sum_{i=1}^q E^{A^i}_0 \;\; ,
\end{equation}
\begin{equation}\label{reference-entropy}
S_0^{A^{\nO{1}}} = \sum_{i=1}^q S^{A^i}_0 \;\; ,
\end{equation}
with the values of $E^{A^i}_0$ and $S^{A^i}_0$ fixed arbitrarily.
Notice that by construction $V_0^{A^{\nO{1}}} = \sum_{i=1}^q
V^{A^i}_0 $ and, therefore, we also have $E_0^{A^{\nO{1}}} +
p_0V_0^{A^{\nO{1}}} = \sum_{i=1}^q (E^{A^i}_0+p_0V^{A^i}_0 )$.  In
chemical thermodynamics, it is customary to set
$E^{A^i}_0+p_0V^{A^i}_0 =0$ and $S^{A^i}_0 = 0$ for each elemental
species.

 Similarly to what
seen for a closed system, the definition of energy for $O$ can be
extended to the states of $O$ in which $O$ is separable but
correlated with its environment.

\section{\label{fundamental-open}Fundamental relation for an open system}

\noindent \textbf{Stable equilibrium state of an open system}. A
state of an open system $O$ in which $O$ is a separable open
system in $\FO$ and is uncorrelated from its environment $Q$ is
called a stable equilibrium state if it can be reproduced as a
stable equilibrium state of a closed system $A$ in $\FA$ = $\FO$.

\spazio

We will consider separately the two different cases:\\ a) the
constituents of $O$ are non-reactive, \emph{i.e.}, no reaction
mechanism is allowed for $O$;\\ b) reactions with stoichiometric
coefficients $\pmb{\nu}^O $ are allowed for $O$.

\spazio

\noindent \textbf{Fundamental relation for the stable equilibrium
states of an open system with non-reactive constituents}. Let
$SE^O$ be the set of all the stable equilibrium states of an open
system $O$ with $r$ non-reactive constituents and $s$ parameters,
$\betaO{}=\beta^O_1$, ... , $\beta^O_s$. Let us consider the
subset $SE^O_{\nO{1}}$ of all the states of $SE^O$ that have the
composition $\nO{1}$, and let $A^{\nO{1}}$ be a closed system with
composition $\nO{1}$, such that its stable equilibrium states
coincide with those of the subset $SE^O_{\nO{1}}$ and therefore
also the parameters coincide, \emph{i.e.},
${\pmb{\beta}^{A^{\nO{1}}}}=\betaO{}$. Then, every subset
$ESE^{A^{\nO{1}}}$ of equivalent stable equilibrium states of
$A^{\nO{1}}$, which is determined by the energy $E^{A^{\nO{1}}}$
and the parameters $\pmb{\beta}^{A^{\nO{1}}}$, coincides with a
subset of equivalent stable equilibrium states of $O$ with
composition $\nO{1}$. The same argument can be repeated for every
composition of $O$. Therefore, on the whole set $SE^O$, a relation
with the form
\begin{equation}\label{fundamental-relation-open}
S_{se}^O = S_{se}^O(E^O, \; \nO{}, \; \betaO{}) \;\;
\end{equation}
is defined and is called fundamental relation for $O$. Since the
relation $S_{se}^O = S_{se}^O (E^O)$, for fixed values of $\nO{}$
and $ \betaO{}$, is strictly increasing, Eq.
(\ref{fundamental-relation-open}) can be rewritten as
 \begin{equation}\label{fundamental-relation-open-two}
E_{se}^O = E_{se}^O(S^O, \; \nO{}, \; \betaO{}) \;\; .
\end{equation}

\spazio

\noindent \textbf{Gibbs equation for a non-reactive open system}.
If the system has non-reactive constituents, the fundamental
relation given by Eq. (\ref{fundamental-relation-open-two})
applies. By differentiating Eq.
(\ref{fundamental-relation-open-two}), one obtains (omitting the
superscript \quot{$O$} and the subscript \quot{se} for simplicity)
\begin{equation}\label{Gibbs-open}
dE = T dS + \sum_{i\,=1}^r \mu_i \; \textrm{d} n_i +
\sum_{j\,=1}^s F_j \; \textrm{d} \beta_j \;\; ,
\end{equation}
where $\mu_i$ is called the \emph{total potential} of $i$-th
constituent of $O$.\\ In Eq. (\ref{Gibbs-open}), it is assumed
that Eq. (\ref{fundamental-relation-open-two}) is continuous and
differentiable also with respect to $\textbf{\emph{n}}$. For
systems with very large values of the amounts of constituents this
condition is fulfilled. However, for very few particle closed
systems, the variable $\textbf{\emph{n}}$ takes on only discrete
values, and, according to our definition, a separable state of an
open system must be reproduced as a separable state of a closed
system. Thus, the extension of Eq. (\ref{Gibbs-open}) to few
particles open systems requires an \emph{extended definition} of a
separable state of an open system, which includes states with non
integer numbers of particles. This extension will not be presented
here.

\spazio

\noindent \textbf{Fundamental relation for the stable equilibrium
states of an open system with reactive constituents}. Let $SE^O$
be the set of all the stable equilibrium states of an open system
$O$ with parameters $\betaO{}$ and constituents which can react
according to a set of reaction mechanisms defined by the
stoichiometric coefficients $\pmb{\nu}^O$. Let $(\pmb{n}^{0O}_1$,
$\pmb{\nu}^O)$ be the set of the compositions of $O$ which are
compatible with the initial composition $\pmb{n}^{0O}_1 =
(n^{0O}_1, ... , n^{0O}_r)_1$. Let $SE^{\pmb{n}^{0O}_1}$ be the
subset of $SE^O$ with compositions compatible with
$(\pmb{n}^{0O}_1$, $\pmb{\nu}^O)$ and let $A^{\pmb{n}^{0O}_1}$ be
a closed system with compositions compatible with
$(\pmb{n}^{0O}_1$, $\pmb{\nu}^O)$ and stable equilibrium states
that coincide with those of the subset $SE^{\pmb{n}^{0O}_1}$ so
that also the parameters coincide, \emph{i.e.},
$\pmb{\beta}^{A^{\pmb{n}^{0O}_1}}= \; \betaO{}$.\\ Then, every
subset $ESE^{A^{\pmb{n}^{0O}_1}}$ of equivalent stable equilibrium
states of $A^{\pmb{n}^{0O}_1}$, which is determined by the energy
$E^{A^{\pmb{n}^{0O}_1}}$ and the parameters
$\pmb{\beta}^{A^{\pmb{n}^{0O}_1}}$, coincides with a subset of
equivalent stable equilibrium states in the set
$SE^{\pmb{n}^{0O}_1}$. The same argument can be repeated for every
set of compatible compositions of $O$, $(\pmb{n}^{0O}_2$,
$\pmb{\nu}^O)$, $(\pmb{n}^{0O}_3$, $\pmb{\nu}^O)$, etc. Therefore,
on the whole set $SE^O$, the following single-valued relation is
defined
\begin{equation}\label{fundamental-relation-open-reactive}
S_{se}^O  = S_{se}^O (E^O, \; \pmb{n}^{0O}, \; \pmb{\beta}^{O})
\;\;
\end{equation}
which is called fundamental relation for $O$. Since the relation
$S_{se}^O = S_{se}^O (E^O)$, for fixed values of $\pmb{n}^{0O}$
and $\pmb{\beta}^{O}$, is strictly increasing, Eq.
(\ref{fundamental-relation-open-reactive}) can be rewritten as
 \begin{equation}\label{fundamental-relation-open-reactive-two}
E_{se}^O = E_{se}^O(S^O, \; \pmb{n}^{0O}, \; \pmb{\beta}^{O}) \;\;
.
\end{equation}

\spazio

\noindent \emph{Comment}. On the set $SE^O$ of the stable
equilibrium states of $O$, also the reaction coordinates are given
by a single valued function
\begin{equation}\label{chem-eq-open-reactive}
\pmb{\varepsilon}_{se}^O  = \pmb{\varepsilon}_{se}^O (E^O, \;
\pmb{n}^{0O}, \; \pmb{\beta}^{O}) \;\; ,
\end{equation}
which defines the chemical equilibrium composition. The existence
of Eq. (\ref{chem-eq-open-reactive}) is a consequence of the
existence of a single valued function such as Eq.
(\ref{chem-eq-composition}) for each of the closed systems
$A^{\pmb{n}^{0O}_1}$, $A^{\pmb{n}^{0O}_2}$, ... used to reproduce
the stable equilibrium states of $O$ with sets of amounts of
constituents compatible with the initial compositions,
$\pmb{n}^{0O}_1$, $\pmb{n}^{0O}_2$, etc.

\section{\label{concl}Conclusions}

In this paper, a general definition of entropy is presented, based
on operative definitions of all the concepts employed in the
treatment, designed to provide a clarifying and useful, complete
and coherent, minimal but general, rigorous logical framework
suitable for unambiguous fundamental discussions on Second Law
implications.

 Operative definitions of system, state, isolated
system, environment of a system, process, separable system, system
uncorrelated from its environment and parameters of a system are
stated, which are valid also in the presence of internal
semipermeable walls and reaction mechanisms. The concepts of heat
and of quasistatic process are never  mentioned, so that the
treatment holds also for nonequilibrium states, both for
macroscopic and few particles systems.

 The role of correlations
on the domain of definition and on the additivity of energy and
entropy is discussed: it is proved that energy is defined for any
separable system, even if correlated with its environment, and is
additive for separable subsystems even if correlated with each
other; entropy is defined only for a separable system uncorrelated
from its environment and is additive only for separable subsystems
uncorrelated from each other; the concept of decorrelation entropy
is defined.

 A definition of thermal reservoir less restrictive
than in previous treatments is adopted: it is fulfilled, with an
excellent approximation, by any single-constituent simple system
contained in a fixed region of space, provided  that the energy
values are restricted to a suitable finite range. The proof that
entropy is a property of the system is completed by a new explicit
proof that the entropy difference between two states of a system
is independent of the initial state of the auxiliary thermal
reservoir chosen to measure it.

 The
definition of a reversible process is given with reference to a
given \emph{scenario}, \emph{i.e.}, the largest isolated system
whose subsystems are available for interaction; thus, the
operativity of the definition is improved and the treatment
becomes compatible also with recent interpretations of
irreversibility in the quantum mechanical framework.

 Rigorous
extensions of the definitions of energy and entropy to open
systems are stated. The existence of a fundamental relation for
the stable equilibrium states of an open system with reactive
constituents is proved rigorously; it is shown that the amounts of
constituents which correspond to given fixed values of the
reaction coordinates should appear in this equation.

\section*{Acknowledgments}

\noindent G.P. Beretta gratefully acknowledges the
Cariplo--UniBS--MIT-MechE faculty exchange program co-sponsored by
UniBS and the CARIPLO Foundation, Italy under grant 2008-2290.

\end{document}